\documentclass[a4paper,11pt]{article}
\pdfoutput=1

\usepackage{jheppub} 
\usepackage{graphicx}
\usepackage{hyperref}
\usepackage[utf8]{inputenc} 
\usepackage{amsmath}
\usepackage{amssymb}
\usepackage{float}
\usepackage{comment}
\usepackage{slashed}
\usepackage[normalem]{ulem}
\usepackage{breqn}
\usepackage{caption}
\usepackage{soul}
\usepackage{subcaption}
\usepackage{booktabs}
\usepackage{mathtools,braket,blkarray}
\usepackage[usenames,dvipsnames]{color}
\usepackage[colorinlistoftodos]{todonotes}
\usepackage[normalem]{ulem}
\usepackage{multirow}
\usepackage[colorinlistoftodos]{todonotes}

 \newcommand {\ignore}[1]{}

\def\21{$\mathrm{SU(2)_L \otimes U(1)_Y}$}

\def\bea{\begin{eqnarray}}
\def\eea{\end{eqnarray}}
\def\beq{\begin{equation}}
\def\eeq{\end{equation}}

\newcommand{\lsim}{
\mathrel{\hbox{\rlap{\hbox{\lower4pt\hbox{$\sim$}}}\hbox{$<$}}}}
\newcommand{\gsim}{
\mathrel{\hbox{\rlap{\hbox{\lower4pt\hbox{$\sim$}}}\hbox{$>$}}}}

\title{\boldmath Terrestrial detection of hidden vectors produced by solar nuclear reactions}

\author[a,b]{Francesco D'Eramo}
\author[c,d]{, Giuseppe Lucente}
\author[d]{, Newton Nath}
\author[a,b,e]{, Seokhoon Yun}

\affiliation[a]{Dipartimento di Fisica e Astronomia, Universit\`a degli Studi di Padova, \\ Via Marzolo 8, 35131 Padova, Italy}
\affiliation[b]{Istituto Nazionale di Fisica Nucleare (INFN), Sezione di Padova, \\ Via Marzolo 8, 35131 Padova, Italy}
\affiliation[c]{Dipartimento Interateneo di Fisica ``Michelangelo Merlin'', Via Amendola 173, 70126 Bari, Italy}
\affiliation[d]{Istituto Nazionale di Fisica Nucleare (INFN), Sezione di Bari, Via Orabona 4, 70126 Bari, Italy}
\affiliation[e]{Center for Theoretical Physics of the Universe, Institute for Basic Science (IBS), Daejeon, 34126, Korea}

\emailAdd{francesco.deramo@pd.infn.it}
\emailAdd{giuseppe.lucente@ba.infn.it}
\emailAdd{newton.nath@ba.infn.it}
\emailAdd{seokhoon.yun@pd.infn.it}

\abstract{
Solar nuclear reactions can occasionally produce sub-MeV elusive beyond the Standard Model particles that escape the solar interior without further interactions. This study focuses on massive spin-one particles. We construct the general theoretical framework and identify two crucial mixing sources involving the photon, which facilitate communication between the hidden and visible sectors: kinetic mixing with the photon, and plasma-induced mixing due to thermal electron loops. For both cases, we focus on the second stage of the solar proton-proton chain and evaluate the fluxes of monochromatic 5.49~MeV hidden vectors produced by the $p(d, ^3{\rm He})\gamma^\prime$ nuclear reaction. We then investigate their terrestrial detection via Compton-like scatterings. The incoming fluxes are polarized, and we evaluate the cross sections for Compton-like scatterings for transverse and longitudinal vectors. Finally, we apply this framework to a concrete case by investigating the sensitivity of the forthcoming Jiangmen Underground Neutrino Observatory (JUNO) experiment and identifying parameter space where current terrestrial bounds will be improved.}

\begin{document} 
\preprint{CTPU-PTC-23-25}
\maketitle
\flushbottom

\section{Introduction}
\label{sec:intro}

The convincing observational evidence for dark matter (DM)~\cite{Bergstrom:2000pn,Bertone:2004pz} leaves no doubt about the need for physics beyond the Standard Model (SM). This is not the only shortcoming of the SM,  an incomplete list of open issues includes neutrino masses, matter-antimatter asymmetry, and the strong CP problem. The presence of a \textit{``dark''} or \textit{``hidden sector"} consisting of light and weakly-coupled particles that do not experience any known interaction naturally appears in motivated frameworks addressing these issues. Examples include the QCD axions~\cite{Peccei:1977hh,Peccei:1977ur,Wilczek:1977pj,Weinberg:1977ma}, dark photons \cite{Okun:1982xi,Galison:1983pa,Holdom:1985ag,Arkani-Hamed:2008hhe}, and new Abelian gauge bosons~\cite{Davidson:1978pm,Mohapatra:1980qe,Fayet:1980ad,Fayet:1980rr,Fayet:1990wx,Wetterich:1981bx,Foot:1990mn,He:1990pn,He:1991qd,Buchmuller:1991ce,Babu:1997st,Fayet:2016nyc}. A noteworthy effort has been recently dedicated to the search for dark sectors~\cite{Jaeckel:2010ni,Essig:2013lka,Alexander:2016aln,Gori:2022vri}, and some discrepancies between theoretical predictions and experimental  measurements have been suggested as possible hints of a dark sector~\cite{Ackerman:2008kmp,Feng:2009mn,Tulin:2013teo,Pospelov:2008zw,TuckerSmith:2010ra,Barger:2010aj,Batell:2011qq}.  These hypothetical degrees of freedom interact with visible matter through several ``portal'' interactions that are constrained by the symmetries of the SM. There are four well-known portals classified in terms of the mediator: Higgs portal, axion portal, neutrino portal, and vector portal.

This work focuses on the phenomenological consequences of a (spin-one) vector portal particle. We assume for concreteness that such a new and light degree of freedom is the gauge boson of additional Abelian gauge symmetry, a scenario that is indeed ubiquitous in many ultraviolet (UV) complete theories such as string compactification for unifying all forces. We also take the dark Abelian gauge symmetry to be broken spontaneously, and the resulting gauge field $\gamma^\prime$ gets a non-vanishing mass. Although this is our general set-up, our analysis can be easily translated into scenarios with a different origin for the vector field (e.g., a composite resonance from a strongly-interacting confining dark sector) with the appropriate identification of couplings. For this reason, we use the term \textit{hidden vector} to refer to the hypothetical particle throughout this study, encompassing scenarios where $\gamma^\prime$ serves as the mediator of a new Abelian force and cases where it does not.

As far as laboratory-based experimental searches are concerned, a high-frequency test of the Coulomb law~\cite{Williams:1971ms,Bartlett:1988yy} and high-quality microwave cavities \cite{Jaeckel:2007ch}~set strong constraints on the mass $m_{\gamma^\prime}$ of hidden vectors. Further constraints arising from the hidden vector decays can be listed as beam-dump~\cite{Bergsma:1985is,Konaka:1986cb,Riordan:1987aw,Bjorken:1988as,Bross:1989mp,Davier:1989wz,Asai:2022zxw},
collider searches~\cite{Aubert:2009cp,Curtin:2013fra,Lees:2014xha,Ablikim:2017aab,Aaij:2017rft,Anastasi:2015qla,Chakraborty:2021apc},
and rare-meson-decay \cite{Bernardi:1985ny,MeijerDrees:1992kd,Archilli:2011zc,Gninenko:2011uv} experiments. Searches using laser experiments are also becoming very important tools to probe dark sector particles \cite{Cameron:1993mr, Gies:2006ca,Ahlers:2006iz,Ahlers:2007rd,Jaeckel:2007gk,Robilliard:2007bq,Chou:2007zz,Ahlers:2007qf}. In addition, stellar objects can be seen as laboratories to study various properties of these light, weakly interacting particles \cite{Raffelt:1996wa,Raffelt:1999tx}. Interestingly, various measurements of the neutrino flux of Supernova SN 1987A~\cite{Kamiokande-II:1987idp,Bionta:1987qt,Alekseev:1987ej} showed how one can set strong constraints on hidden-sector particles like axions~\cite{Turner:1987by,Raffelt:1987yt,Fischer:2016cyd,Chang:2018rso,Carenza:2019pxu,Croon:2020lrf,Lucente:2020whw,Fischer:2021jfm,Caputo:2021rux,Choi:2021ign,Lella:2022uwi} and dark gauge bosons~\cite{Dent:2012mx,An:2014twa,Chang:2016ntp,Hardy:2016kme,Knapen:2017xzo,Hong:2020bxo,Shin:2021bvz,Shin:2022ulh}. The Sun itself is one of the most important natural sources of feebly interacting particles that could be detected on Earth~\cite{Sikivie:1983ip}. An attempt to analyze the axion flux from nuclear processes in the Sun has been made in Ref.~\cite{Raffelt:1982dr}. Several helioscope experiments, including CAST~~\cite{CAST:2009klq,CAST:2009jdc}, come up with their sensitivity by aiming to detect solar axions, while the latest SNO data~\cite{Bhusal:2020bvx} provide a bound on the axion-nucleon coupling. Borexino has provided bounds on different axion couplings~\cite{Borexino:2012guz} by considering solar axion produced by the $p + d \rightarrow{\rm{^3He}}+ a (5.49 \,\  {\rm MeV})$ nuclear reaction~\cite{Weinberg:1977ma}. A similar analysis has been performed recently by Ref.~\cite{Lucente:2022esm} to improve the Borexino bounds with the next-generation  Jiangmen Underground Neutrino Observatory (JUNO). 

In this work, we continue with the same approach developed in \cite{Lucente:2022esm} and extend the idea to investigate hidden vectors $\gamma^\prime$ produced by the $ p \, + \,  d \,  \rightarrow \, ^3{\rm He} \, + \, \gamma^{\prime} \left(5.49\,{\rm MeV}\right)$ nuclear reaction. Given the set of future detectors, the planned JUNO provides an ideal platform to search for monochromatic and weakly interacting particles, since it combines a large fiducial mass ($\sim20$ kton) and an exquisite energy resolution (3\%/$\sqrt{E({\rm MeV}}$)~\cite{JUNO:2015zny}. 

The landscape of UV completions for the general set-up analyzed here is rather broad. We describe the general theoretical framework in Sec.~\ref{sec:vectors}, where we do not specify the dark gauge charges of the SM fields. Before doing so, we discuss the identification of physical states in the most general case and argue how this procedure is different in dense media (such as the solar core when production occurs) and in the vacuum (approximately the detector environment). By identifying the physical fields, we can uncover the key mixing mechanisms responsible for the concealed hidden vector interactions with SM particles. On the one hand, a renormalizable kinetic mixing between the two Abelian field strengths is allowed by the symmetries of the theory, and it would be enough to ensure viable production and detection channels. Even if the kinetic mixing is absent, we can still have a consistent story of production in the solar core followed by terrestrial detection thanks to plasma mixing effects as long as the electron carries a vector-like dark gauge charge. Electron thermal loops generate an effective mixing between the hidden vector and the SM photon responsible for the production, whereas detection goes on regularly via electron coupling. 

Having these two mixing sources in mind, Sec.~\ref{sec:prodet} identifies the broad framework studied in this work. Our focus is on the Abelian extension of the SM gauge group where dark charges of SM fields, if any, are vector-like. Furthermore, we require that dark gauge charges are consistent with electroweak gauge invariance. We perform a complete analysis where we provide production and detection rates in Sec.~\ref{sec:prodet}, and JUNO projections in Sec.~\ref{sec:JUNOdet}. Throughout our work, we identify the two main new physics scenarios described below.
\begin{itemize}
    \item \textbf{Kinetic Mixing.} The only communication channel between visible particles and the hidden vector is a non-vanishing kinetic mixing. Production rates for transverse and longitudinal polarizations are provided in Fig.~\ref{fig:DPflux}, and the JUNO sensitivity on the relevant parameter space is shown in Fig.~\ref{Fig:JUNO_DP}.
    \item \textbf{Plasma Mixing.} The kinetic mixing is assumed absent or subdominant and the effect is controlled by a non-vanishing vector-like dark gauge charge of the electron field. Plasma effects induce an effective $\gamma / \gamma^\prime$ mixing that is responsible for the production rate provided in Fig.~\ref{fig:BLflux}, and the resulting JUNO projections in Fig ~\ref{Fig:JUNO_IM}.
\end{itemize}

Broadly speaking, current experimental bounds as well as projections for future experiments can be classified into two main categories. On the one hand, the hidden vector can be stable on cosmological timescales and present in the Universe with a relic density that reproduces the observed one for dark matter. If this is the case, experimental searches looking for vector dark matter proceed with the implicit assumption that there is a local mass density $\rho_{\gamma^\prime}^{\odot} \simeq 0.4 \, {\rm GeV} / {\rm cm}^3$ in the solar system. However, the hidden vector could still exist without being stable and/or cosmologically abundant. Even in this case, there is plenty of experimental signatures due to these new particles. Solar production is one of the most studied examples, and there are two main strategies to search for them: looking for deviations in the stellar properties (e.g., cooling), and direct detection of these hidden vectors here on Earth. The latter approach, which is the one pursued in this work with direct detection at JUNO, leads typically to weaker bounds since there is a higher price to pay in terms of feeble couplings to the hidden sector. This is confirmed by Figs.~\ref{Fig:JUNO_DP} and \ref{Fig:JUNO_IM}. Remarkably, Fig.~\ref{Fig:JUNO_DP} shows how JUNO will provide the best terrestrial bounds in the mass range $50 \, {\rm keV} \lesssim m_{\gamma^\prime} \lesssim  1 \, {\rm MeV}$ for the kinetic mixing scenario.

We conclude our study in Sec.~\ref{sec:Conclusion}. As illustrated by Figs.~\ref{fig:DPflux} and \ref{fig:BLflux}, the Sun provides a polarized flux of hidden vectors, and the Compton-like scattering cross section for detection has to be evaluated accordingly for the two different polarizations of the incoming particle. We spelled out the technical details of these calculations in App.~\ref{app:Compton}.

\section{Massive Hidden Vectors: General Framework}
\label{sec:vectors}

Our general analysis is valid for any massive spin-one particle coupled to light SM degrees of freedom. It is reasonable to imagine a microscopic theory in which this new hypothetical degree of freedom interacts with all SM particles. However, when considering the low energies of interest, a phenomenological approach becomes viable, focusing only on a few key interactions that are relevant. In this section, we briefly comment on the UV origin of our framework and then we develop the low-energy Lagrangian needed for our analysis. Finally, we identify the physical states arising from different mixing effects and derive their interactions. Crucially, this procedure in the vacuum is different from the same operation in the solar interior where the large electron density leads to collective effects that alter particle propagators. For this reason, we discuss these two cases separately. 

Our conceptual starting point is a UV complete theory valid at energies high enough so that the SM enjoys the full $\mathcal{G}_{\rm SM} = SU(3)_c \times SU(2)_L \times U(1)_Y$ gauge symmetry. We extend the SM field content with a new massive spin-one particle $\gamma^\prime$, and we denote by $A_\mu^\prime$ the associated quantum field. For concreteness, we focus on the case where $A_\mu^\prime$ is an Abelian gauge boson of a new local symmetry $U(1)^\prime$ and therefore the full gauge symmetry in the complete  theory is $\mathcal{G}_{\rm full} = \mathcal{G}_{\rm SM} \times U(1)^\prime$. However, this working hypothesis is not necessary and our phenomenological analysis captures also other scenarios such as the one of a massive vector resonance arising from strong dynamics in the dark sector. We remain agnostic about the origin of the vector boson mass $m_{\gamma^\prime}$, and we do not commit to any specific scenario. Examples include just adding a Stueckelberg mass or a spontaneous breaking of the $U(1)^\prime$ via a dark Higgs mechanism. While the Stueckelberg case contains a sufficiently heavy scalar field for $\gamma^\prime$ to acquire a mass, the Higgs mechanism case introduces an additional Higgs field, the excitation of which could affect low-energy phenomenology~\cite{An:2013yua}. In this paper, we consider such a scalar particle heavy enough to be integrated out, thus phenomenological consequences of $\gamma^\prime$ do not rely on how to realize its mass. In general, our implicit assumption is that whatever mechanism provides such a mass does not affect physics at the low energies under our investigation.

If one restricts to renormalizable interactions, there are two main channels for the vector field $A_\mu^\prime$ to communicate with the visible universe. First, its field strength can kinetically mix with the SM Abelian counterpart. At high energies, where the electroweak symmetry is unbroken, the only gauge-invariant kinetic mixing is with the hypercharge. Once we consider energies below the weak scale, with the electroweak group broken into the electromagnetic subgroup, the field $A_\mu^\prime$ kinetically mixes with both the photon $A_\mu$ and the massive $Z_\mu$ boson. The latter interaction leads to negligible effects at low energies and we ignore it. The second option is realized concretely when SM matter fields carry non-vanishing $U(1)^\prime$ charges. The lack of gauge anomalies, perhaps with the aid of dark sector fermion fields, is a non-trivial requirement that the theory must satisfy in this case~\cite{Carone:1994aa,Appelquist:2002mw,Dudas:2013sia,FileviezPerez:2014lnj,Dobrescu:2014fca,Okada:2016gsh,Ismail:2016tod,Okada:2016tci,Cui:2017juz,Bauer:2018egk,Costa:2019zzy,FileviezPerez:2019jju}. 

At energies below the weak scale, we can integrate out the weak gauge bosons, the top quark, and the Higgs boson. The residual unbroken gauge group is $SU(3)_c \times U(1)_{\rm EM} \times U(1)^\prime$, and the Lagrangian of the framework we have just described takes the form
\bea
\mathcal{L} = -\frac{1}{4}F_{\mu\nu}F^{\mu\nu} - \frac{1}{4} F_{\mu\nu}^\prime F^{\prime \mu\nu} + \frac{m_{\gamma^\prime}^2}{2}A_\mu^\prime A^{\prime \mu} +  \frac{\varepsilon}{2}F_{\mu\nu}F^{\prime \mu\nu} + e J^\mu_{\rm EM} A_\mu + e^\prime  J^{\prime \mu} A^\prime_\mu \ .
\label{eq:LSMApv1}
\eea
We omit kinetic and mass terms for fermion fields, and we do not specify QCD interactions since we always work in the phase where nuclear matter is confined. We ignore higher-dimensional operators suppressed by powers of the weak scale. The first three operators in Eq.~\eqref{eq:LSMApv1} are quadratic in the Abelian gauge fields and diagonal.
The kinetic terms contain the field strengths $F_{\mu\nu} = \partial_\mu A_\nu - \partial_\nu A_\mu$ and $F^\prime_{\mu\nu} = \partial_\mu A^\prime_\nu - \partial_\nu A^\prime_\mu$, and we account for a non-vanishing mass $m_{\gamma^\prime}$ for the dark gauge boson. We define $e$ and $e^\prime$ to be the gauge coupling constants for $U(1)_{\rm EM}$ and $U(1)^\prime$, respectively. Photon interactions respect electromagnetic gauge invariance, and they are proportional to the electromagnetic current $J^{\mu}_{\rm EM} = \sum_\psi q_\psi \bar{\psi}\gamma^\mu \psi$, where the sum runs over all SM fermions $\psi$ (with electric charge $q_\psi$). The next paragraph describes the two operators mixing dark and visible sectors.

A renormalizable kinetic mixing between the two Abelian factors proportional to a dimensionless coefficient $\varepsilon$ is allowed by gauge invariance. Even if it is vanishing at some high scale $\Lambda_{\rm UV}$, it can be induced radiatively at low energies by heavy matter fields that carry both Abelian charges. A Dirac fermion $\Psi$ that carries electric charge $q_\Psi$ and dark gauge charge $q_\Psi^\prime$\footnote{We take vector-like quantum numbers for $\Psi$ so that it couples to $A^\prime_\mu$ with the vector current.} induces a kinetic mixing term at one-loop level, which at low energy is given by~\cite{Holdom:1985ag}
\bea
\varepsilon = \varepsilon\left(\Lambda_{\rm UV}\right) + \sum_\Psi N_\Psi \frac{(e  q_\Psi) (e^\prime q_\Psi^\prime)}{12\pi^2} \log\left(\frac{\Lambda_{\rm UV}^2}{m_\Psi^2}\right) \ , 
\eea
where $N_\Psi$ denotes the number of its internal degrees of freedom (e.g., spin, color). The scenario in which the only communication channel is a non-zero kinetic mixing has been commonly referred to as the \textit{dark photon} since the associated physical eigenstate in the vacuum couples to the SM electromagnetic current, which will be explained in Sec.~\ref{subsec:IntVacuum}. For SM fields carrying dark gauge charges, we have the last term in Eq.~\eqref{eq:LSMApv1}. We keep the discussion general for this section without specifying the explicit form of $J^{\prime \mu}$. At high energy, and in the general case, we have dark gauge boson interactions with both vector and axial currents of quarks and leptons. At the low energies that are relevant for production and detection, we have to translate these couplings into interactions with baryons. 

We specify the explicit expression for $J^{\prime \mu}$ in the next section and compute explicitly the rates. Before doing so, we have to switch to a canonical field basis and identify physical states. This procedure is different in the vacuum and in the solar interior where the hot plasma affects vector self-energies. Thus, we discuss the two cases separately.

\subsection{Physical states I: vacuum}
\label{subsec:IntVacuum}

Once massive vectors reach terrestrial detectors, the environmental conditions allow for the use of Eq.~\eqref{eq:LSMApv1} without further corrections. We identify the canonical basis via 
\bea
A_\mu \; \rightarrow \; A_\mu + \frac{\varepsilon}{\sqrt{1-\varepsilon^2}} A^\prime_\mu \ , \qquad \qquad \qquad \qquad
A^\prime_\mu \; \rightarrow \; \frac{1}{\sqrt{1-\varepsilon^2}} A^\prime_\mu \ .
\eea
This $GL(2,R)$ transformation~\cite{Babu:1997st} leads to canonical kinetic terms for any finite value of $\varepsilon$. Phenomenological constraints impose $\varepsilon \ll 1$ and we neglect $\mathcal{O}(\varepsilon^2)$ terms. Thus, the vector mass $m_{\gamma^\prime}$ is unaffected by this transformation. Gauge boson interactions with fermions in the canonical basis and at the leading order in $\varepsilon$ result in
\bea
\left. \mathcal{L} \right|_{\rm vacuum} \supset  e J^\mu_{\rm EM} A_\mu + \left(\varepsilon e J^\mu_{\rm EM} + e^\prime  J^{\prime \mu}\right) A^\prime_\mu \ .
\label{eq:DPeffVac}
\eea
Electromagnetic interactions remain the same as in the SM, whereas the dark gauge boson acquires an additional $\varepsilon$-suppressed coupling to the electromagnetic current. 

\subsection{Physical states II: medium}

The procedure to identify canonically normalized fields is different in the solar interior because the hot ionized plasma affects particle self-energies. This effect is already present in the SM case with a modification of the photon self-energy $\Pi^{\mu\nu}_{\gamma\gamma}$~\cite{Altherr:1992jg,Braaten:1993jw}. Another important effect for our analysis is an effective mass mixing $\Pi_{\gamma\gamma^\prime}^{\mu\nu}$ between the two vectors~\cite{Redondo:2008ec,An:2013yfc,Redondo:2013lna,Hardy:2016kme}. There is also a correction to the diagonal self-energy $\Pi_{\gamma^\prime \gamma^\prime}^{\mu\nu}$, but it can be neglected since there is a higher price to pay in terms of couplings mixing the two sectors. We conclude this section with proper treatment of medium effects, and we provide the effective Lagrangian needed to evaluate solar production. 

Medium effects alter the transverse photon polarizations and provide a longitudinal one. For a photon with four-momentum $K^\mu = (\omega, \vec{k})$, we have the polarization vectors
\begin{align}
\label{eq:epsilonT} \text{Transverse: } & \; \; \epsilon_{\rm T}^\mu = 
\left(0, \frac{\hat{e}_1 \pm i \hat{e}_2}{\sqrt{2}} \right) \ , \\
\label{eq:epsilonL} \text{Longitudinal: } & \; \; \epsilon_{\rm L}^\mu = 
\frac{1}{\sqrt{\omega^2-k^2}} \left(k, \omega \hat{k} \right) \ .
\end{align}
Here, $\hat{e}_1$ and $\hat{e}_2$ are two normal unit-vectors in the plane orthogonal to the propagation direction, and we define the magnitude of the spatial momentum $k=|\vec{k}|$ and its direction $\hat{k} = \vec{k} / k$. The polarization tensor reads as
\bea
\Pi^{\mu\nu}_{\gamma\gamma} = \left<(e J_{\rm EM}^\mu) \, (e J_{\rm EM}^\nu)\right> =
\pi_{\rm T}\sum_{\rm T = +,-} \epsilon_{\rm T}^{\mu}\epsilon_{\rm T}^{*\nu} + \pi_{\rm L} \, \epsilon_{\rm L}^{\mu}\epsilon_{\rm L}^{*\nu} \, .
\eea
The two factors $\pi_{\rm T}$ and $\pi_{\rm L}$ account for medium effects. The real parts of $\pi_{\rm T,L}$ modify the dispersion relations for the respective polarizations, whereas the imaginary parts of $\pi_{\rm T,L}$ are associated with the absorption of the excitation with four-momentum $K^\mu = (\omega, \vec{k})$~\cite{Weldon:1983jn}. For our purposes, we can take the limit where medium electrons are non-degenerate and non-relativistic, and in this case, we have the simple expressions~\cite{Redondo:2008aa,An:2013yfc} 
\bea
{\rm Re}\left[\pi_{\rm T}\right]  \simeq \omega_{\rm pl}^2  \, , \qquad \qquad
{\rm Re}\left[\pi_{\rm L}\right] \simeq \omega_{\rm pl}^2 \left(1 - \frac{k^2}{\omega^2} \right) \ .
\label{eq:RePi}
\eea
The plasma frequency $\omega_{\rm pl}$ is given by
\bea
\omega_{\rm pl} = \sqrt{\frac{4\pi \alpha n_e}{m_e}} = 0.30\,{\rm keV}\left(\frac{n_e}{6.3 \times 10^{25} \, {\rm cm}^{-3}}\right)^{1/2}\, ,
\label{eq:PlasmaFreq}
\eea
where, $m_e$ denotes the electron mass, $n_e$ represents the electron density, and $\alpha$ corresponds to the fine structure constant. The last expression is evaluated for typical electron densities in the solar core. A medium-induced dispersion is typically momentum dependent due to the breaking of Lorentz invariance by singling out the medium rest frame. The imaginary parts of $\pi_{\rm T,L}$ become relevant once we approach the resonant region $\omega^2 - k^2 = {\rm Re}\left[\pi_{\rm T,L}\right]$. In this work, the typical energies at play are large, $\omega \simeq 5.49\,{\rm MeV} \gg \omega_{\rm pl} \simeq 0.3\,{\rm keV}$.
Hence, while the resonance condition cannot be achieved for the longitudinal component, the production of the transverse ones would be resonantly enhanced at $\omega^2 - k^2 = {\rm Re}[\pi_{\rm T}] = \omega_{\rm pl}^2$.
The compton scattering controls the absorption rate in the high energy regime~\cite{Redondo:2008aa,An:2013yfc}, and the imaginary part of $\pi_{\rm T}$ reads as
\bea
{\rm Im}\left[\pi_{\rm T}\right] \simeq -\omega\left(1-e^{-\omega/T}\right)\frac{8\pi \alpha^2}{3m_e^2}n_e\, .
\label{eq:ImPiT}
\eea

Plasma effects provide an additional mixing source through thermal loops of particles carrying both electromagnetic and $U(1)^\prime$ charges. This effect is controlled by the lightest particles in the plasma, and we consider the case where electrons are charged under both the Abelian gauge symmetries since the effect of baryons is suppressed by their heavy mass. The term induced by an electron loop is given by~\cite{Hardy:2016kme,Hong:2020bxo,Shin:2021bvz,Shin:2022ulh}
\bea
\Pi^{\mu\nu}_{\gamma\gamma^\prime} = \left<(e J_{\rm EM}^\mu) (e^\prime J^{\prime\nu})\right> = \frac{e^\prime q_e^\prime}{eq_e}\Pi_{\gamma\gamma}^{\mu\nu} \, ,
\label{eq:Piggp}
\eea
where $q_e (=-1)$ and $q_e^\prime$ denote the vector-like dark charge of the electron.

\begin{figure}[t!]
\centering
    \begin{subfigure}{0.32\textwidth}
    \centering
        \includegraphics[scale=.7]{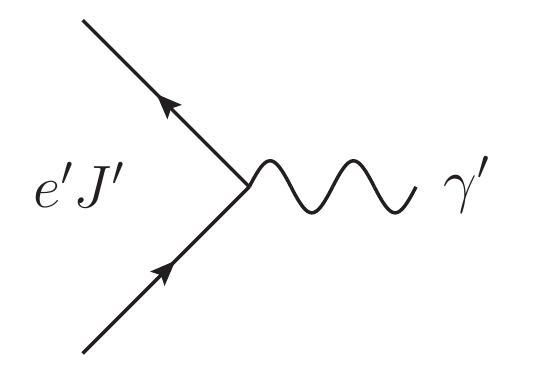}
        \caption{$U^\prime(1)$ charges}
        \label{fig:darkcharge}
    \end{subfigure}
    \hfill
    \begin{subfigure}{0.32\textwidth}
        \includegraphics[scale=.7]{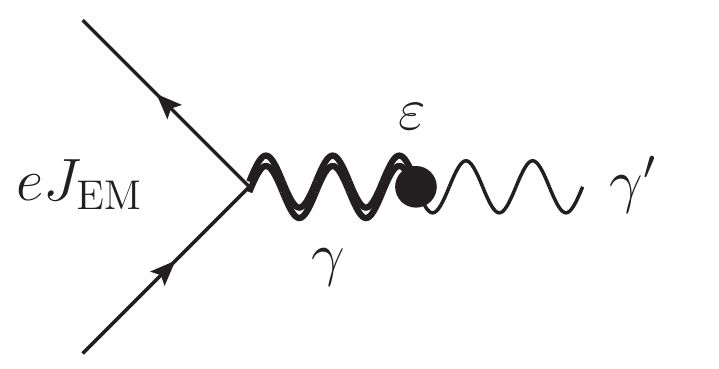}
        \caption{the kinetic mixing $\varepsilon$}
        \label{fig:kineticM}
    \end{subfigure}
    \hfill
    \begin{subfigure}{0.32\textwidth}
        \includegraphics[scale=.7]{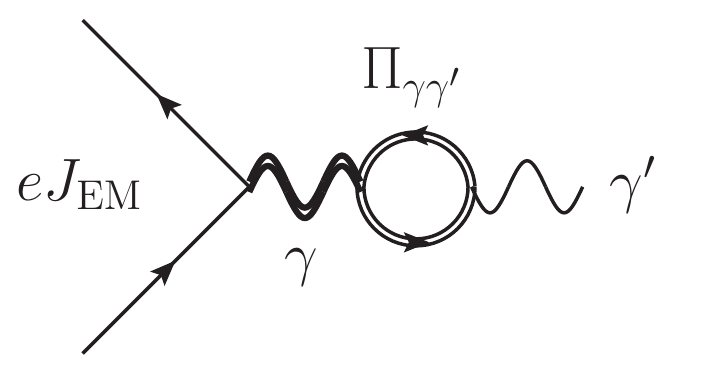}
        \caption{the plasma mixing $\Pi_{\gamma\gamma^\prime}$}
        \label{fig:inMediumM}
    \end{subfigure}
\caption{\footnotesize Feynman diagrams responsible for the effective dark gauge boson interactions in the medium.}
\label{fig:EffectiveCoupling}
\end{figure}
 
The diagrams in Fig.~\ref{fig:EffectiveCoupling} illustrate the effective $\gamma^\prime$ interactions in the solar interior. A tree-level coupling proportional to the dark gauge coupling $e^\prime$ could be there if the dark current $J^{\prime \mu}$ is non-vanishing (Fig.~\ref{fig:darkcharge}). If a non-vanishing kinetic mixing couples the vector field $A_\mu^\prime$ to the electromagnetic current, one has to include the thermally resummed photon propagator as denoted by the double wavy line in Fig.~\ref{fig:kineticM}. Finally, the quadratic mixing in Eq.~\eqref{eq:Piggp} along with the same resummed photon propagator leads to an additional interaction between the dark gauge boson and the electromagnetic current. The loop with the double line in Fig.~\ref{fig:inMediumM} indicates the leading effect leading of the correction $\Pi_{\gamma\gamma^\prime}^{\mu\nu}$ given in Eq.~\eqref{eq:Piggp}. These three interactions are described in the weak mixing regime ($\varepsilon, e^\prime \ll 1$) by the following effective Lagrangian
\bea
\left. \mathcal{L} \right|_{\rm medium} \supset e^\prime J^{\prime \mu} A_\mu^\prime + \left(\varepsilon \frac{m_{\gamma^\prime}^2}{m_{\gamma^\prime}^2 - \pi_{\rm T,L}}  + \frac{e^\prime q_e^\prime}{eq_e} \frac{\pi_{\rm T,L}}{m_{\gamma^\prime}^2-\pi_{\rm T,L}}\right) e J_{\rm EM}^\mu A_{\mu}^\prime \, .
\label{eq:DPeffMed}
\eea
We always consider on-shell (i.e., $\omega^2 - k^2 = m_{\gamma^\prime}^2$) hidden vectors on external legs. When medium effects are absent (i.e., $\pi_{\rm T,L}=0$), Eq.~\eqref{eq:DPeffMed} correctly recovers the expression valid in the vacuum as given in Eq.~\eqref{eq:DPeffVac}.

\section{Production and Detection Rates}
\label{sec:prodet}

Having set up the general formalism, we are now ready to evaluate production and detection rates. As a first step, we specify an explicit realization of the framework described by Eq.~\eqref{eq:LSMApv1} focusing on theories where dark gauge charges of SM fermions, if any, are vector-like. The current coupled to $A_\mu^\prime$ contains only the vector part (i.e., $J^{\prime \mu} = \sum_\psi q_\psi^\prime \bar{\psi} \, \gamma^\mu \psi$), thus Ward-Takahashi identity~\cite{Ward:1950xp,Takahashi:1957xn} holds $\partial_\mu J^{\prime \mu}=0$ and the current $J^{\prime \mu}$ is conserved at the classical level for this class of theories. This conservation law could be broken by gauge anomalies (e.g., baryon or lepton numbers), and this would lead to energy-enhanced production rates and significant constraints from high-energy experiments~\cite{Dror:2017nsg,Ismail:2017ulg}.

In order to deal with solar production and terrestrial detection, we need to identify in the Lagrangian in Eq.~\eqref{eq:LSMApv1} the interactions with the only phenomenologically relevant degrees of freedom: proton ($p$), neutron ($n$), electron ($e$), and neutrinos ($\nu_i$). Thus the low-energy Lagrangian needed for the phenomenology we are interested in reads as
\begin{equation}
\begin{split}
\mathcal{L}_{\rm low-energy} = & \, -\frac{1}{4}F_{\mu\nu}F^{\mu\nu} - \frac{1}{4} F_{\mu\nu}^\prime F^{\prime \mu\nu} + \frac{m_{\gamma^\prime}^2}{2}A_\mu^\prime A^{\prime \mu} +  \frac{\varepsilon}{2}F_{\mu\nu}F^{\prime \mu\nu} + \\ &  \;\; \; e \left( \bar{p} \gamma^\mu p - \bar{e} \gamma^\mu e \right) A_\mu 
+ \\ & \;\; \; e^\prime \left(q^\prime_p \bar{p} \gamma^\mu p + q^\prime_n \bar{n} \gamma^\mu n + q^\prime_e \bar{e} \gamma^\mu e + \sum_i q^\prime_{\nu_i} \bar{\nu_i} \gamma^\mu P_L \nu_i \right) A^\prime_\mu \ . 
\label{eq:LSMApv2}
\end{split}
\end{equation}
Interactions with the photon $A_\mu$ are proportional to the electromagnetic current, and baryon interactions with the dark gauge boson are quantified by the following couplings
\bea
q_p^\prime \equiv 2 q_u^\prime + q_d^\prime \ , \qquad \qquad 
q_n^\prime \equiv & q_u^\prime + 2 q_d^\prime \ .
\label{eq:qNvsqq}
\eea
Electron and neutrino couplings are related by electroweak gauge invariance ($q^\prime_{e_i} = q^\prime_{\nu_i}$ with $e_i$ the charged leptons), and the projector $P_L = (1 - \gamma^5) / 2$ isolates the appropriate neutrino chirality. The Lagrangian in Eq.~\eqref{eq:LSMApv2} contains just a few parameters: vector boson mass, kinetic mixing, and three dimensionless couplings (neutrino couplings follow from gauge invariance). Any UV complete model can be mapped into this low-energy effective theory. 

It is useful to discuss the consequences of electroweak gauge invariance on the isospin properties of the low-energy interactions in Eq.~\eqref{eq:LSMApv2}. Isovector couplings originate from tree-level isospin breaking term if $q^\prime_p \neq q^\prime_n$. The relations in Eq.~\eqref{eq:qNvsqq} imply that this is possible if we have $q^\prime_u \neq q^\prime_d$, and such a charge assignment with the same component of a weak doublet having different Abelian charges would not respect $SU(2)_L$ gauge invariance. Thus we constrain $U(1)^\prime$ charge assignments to be isoscalar. As an example, the baryon number $B$ would fit within our choice. 

\subsection{Production from solar nuclear reactions}
\label{sec:production}

Hidden vectors can be produced non-thermally as final states of solar nuclear reactions. We focus on the second step of the proton-proton chain, $p \, + \, d \,  \rightarrow \, ^3{\rm He} \, + \, \gamma \left(5.49\,{\rm MeV}\right)$, where protons and deuteriums form helium nuclei~\cite{Raffelt:1982dr}. While this nuclear process produces a photon $\gamma$ most of the time, it can lead occasionally to a monochromatic $\gamma^\prime$ via the mixing effects discussed in the previous section. The final state hidden vector would have the same energy of $5.49\,{\rm MeV}$, and the production is controlled by its interactions with nucleon fields.

At the leading order in the multipole expansion~\cite{Goldhaber:1955ya,De-Shalit:1967fpa,Donnelly:1975ze}, deuterium fusion proceeds via electric dipole ``E1" and magnetic dipole ``M1" transitions that are characterized by orbital angular momentum $l=1$ (p-wave) and $l=0$ (s-wave), respectively. Both transitions in the long wavelength region are governed by the isovector component of the interactions with nucleons. For the E1 transition, this follows from the fact that the proton and the deuteron are pure isospin eigenstates and  there is no electric dipole radiation when the center-of-mass frame coincides with the center-of-charge frame~\cite{Goldhaber:1955ya,Eichmann:1963,De-Shalit:1967fpa}. The magnetic dipole transition is also predominantly isovector~\cite{Raffelt:1982dr,CAST:2009klq,Massarczyk:2021dje}. 

Hidden vectors cannot be produced via the tree-level interaction illustrated by Fig.~\ref{fig:darkcharge} because we choose pure isoscalar couplings (i.e., $q^\prime_u = q^\prime_d$ leading to $q^\prime_p = q^\prime_n$). Providentially, a non-vanishing isovector component is effectively generated by the mixing with the photon because electromagnetic interactions violate isospin due to the different electric charges of up and down quarks. As illustrated in Figs.~\ref{fig:kineticM} and \ref{fig:inMediumM}, and quantified explicitly in Eq.~\eqref{eq:DPeffMed}, there are two mixing sources: kinetic mixing ($\varepsilon \neq 0$) and plasma mixing ($q_e^\prime \neq 0$). They both lead to isovector interactions with nucleons~\cite{Shin:2021bvz,Shin:2022ulh,Chun:2022qcg}. 

The fluxes provided in this subsection need to be understood as the number of vectors produced in the solar core per unit of time divided by the surface of a sphere with a radius equal to one astronomical unit. This does not need to be the actual flux hitting a detector on Earth since attenuation processes can be in action between production and detection. We quantify their importance in the next subsection where we discuss absorption from the solar medium and decays outside the Sun during the propagation towards the Earth.

We examine below the two distinct scenarios where only one mixing source is switched on. This applies to any generic $U(1)^\prime$ extension of the SM containing both sources since one of them typically dominates the $\gamma^\prime$ production rate.

\subsubsection*{Production via kinetic mixing}

The scenario where the communication with the SM proceeds via the kinetic mixing only has been dubbed in the literature as the \textit{dark photon}. In this case, the interaction responsible for production in the solar core is the first term in the parenthesis of Eq.~\eqref{eq:DPeffMed}. Dark photon interactions with the SM particles are proportional to the EM current, and the production of $\gamma^\prime$ particles via the deuterium fusion simply follows the photon one with an additional factor involving the plasma effect as well as the kinetic mixing parameter $\varepsilon$. The flux of $5.49\,{\rm MeV}$ transverse (T) and longitudinal (L) dark photons results in~\cite{Donnelly:1978ty,Avignone:1988bv,Haxton:1991pu}
\begin{align}
\label{eq:DPFluxT} \left[\Phi_{\gamma^\prime 0}\right]_{\rm T}^{\varepsilon} \simeq & \, 
\frac{\phi_\gamma}{4\pi d_{\odot}^2} \times \frac{k^3}{\omega^3} \times 
\left|\varepsilon \frac{m_{\gamma^\prime}^2}{m_{\gamma^\prime}^2-\pi_{\rm T}}\right|^2 \ , \\
\label{eq:DPFluxL} \left[\Phi_{\gamma^\prime 0}\right]_{\rm L}^{\varepsilon} \simeq & \,
\frac{\phi_\gamma}{4\pi d_{\odot}^2} \times \frac{k^3}{\omega^3} \times \left|\varepsilon   \frac{m_{\gamma^\prime}^2}{m_{\gamma^\prime}^2-\pi_{\rm L}}\right|^2 \times \frac{1-P_{\rm M1}}{2} \times \frac{m_{\gamma^\prime}^2}{\omega^2} \ .
\end{align}
The subscript `0' in the above expressions denotes the flux on Earth if it is not diminished by effects such as solar medium absorption or vector decays. Both expressions are proportional to the deuterium fusion rate $\phi_{\gamma} = 1.7 \times 10^{38}\,{\rm s}^{-1}$~\cite{Bahcall:1972ax,Raffelt:1982dr,Serenelli:2011py,Borexino:2017rsf} and normalized per unit surface $4 \pi d_{\odot}^2$ with $d_{\odot} = 1.5\times 10^{11}\,{\rm m}$ the Earth-Sun distance. The modified dispersion relation for a massive spin-one particle, namely $k \neq \omega$ since the spatial momentum of the hidden vector is related to its energy $\omega$ via $k = \sqrt{\omega^2 - m_{\gamma^\prime}^2}$, brings three additional powers of the ratio $(k/\omega)$; one power is just due to the final state phase-space, and the two remaining ones follow from the multipole expansion~\cite{Donnelly:1978ty}. Both relations receive an additional multiplicative correction proportional to the kinetic mixing parameter squared and the resummed thermal photon propagator with self-energies $\pi_{\rm T,L}$ given by Eqs.~\eqref{eq:RePi} and \eqref{eq:ImPiT}. There are no additional corrections for the production rate of transverse dark photons. For the longitudinal polarization, there are two additional corrections that we have to account for. First, we have to isolate the contribution due to the E1 transition because longitudinal polarizations are intrinsically electric (i.e., $\vec{k} \times \vec{\epsilon}_{\rm L}=0$), and we also have to put an additional overall $1/2$ factor  for the longitudinal degree of freedom compared with the transverse components. This is the origin of the $(1 - P_{\rm M1})/2$ factor with ${P_{\rm M1}= 54 \%}$~\cite{Schmid:1997zz} the fraction of the M1 transition at the relevant energy. Second, we also have an additional multiplicative factor $(m_{\gamma^\prime}/\omega)^2$ that follows from the conservation law of the electromagnetic current.\footnote{The transition amplitude is proportional to the electromagnetic current which is conserved $k_\mu J^\mu_{\rm EM} = \omega J^0_{\rm EM} - k J^3_{\rm EM} = 0$. Thus the amplitude is proportional to the factor $\epsilon_{{\rm L} ,\mu}^* J^\mu_{\rm EM} = \frac{k J^0_{\rm EM} - \omega J^3_{\rm EM}}{m_{\gamma^\prime}} = - J^3_{\rm EM} \frac{m_{\gamma^\prime}}{\omega}$.} 

\begin{figure}[t!]
\centering
\includegraphics[width=0.7\textwidth]{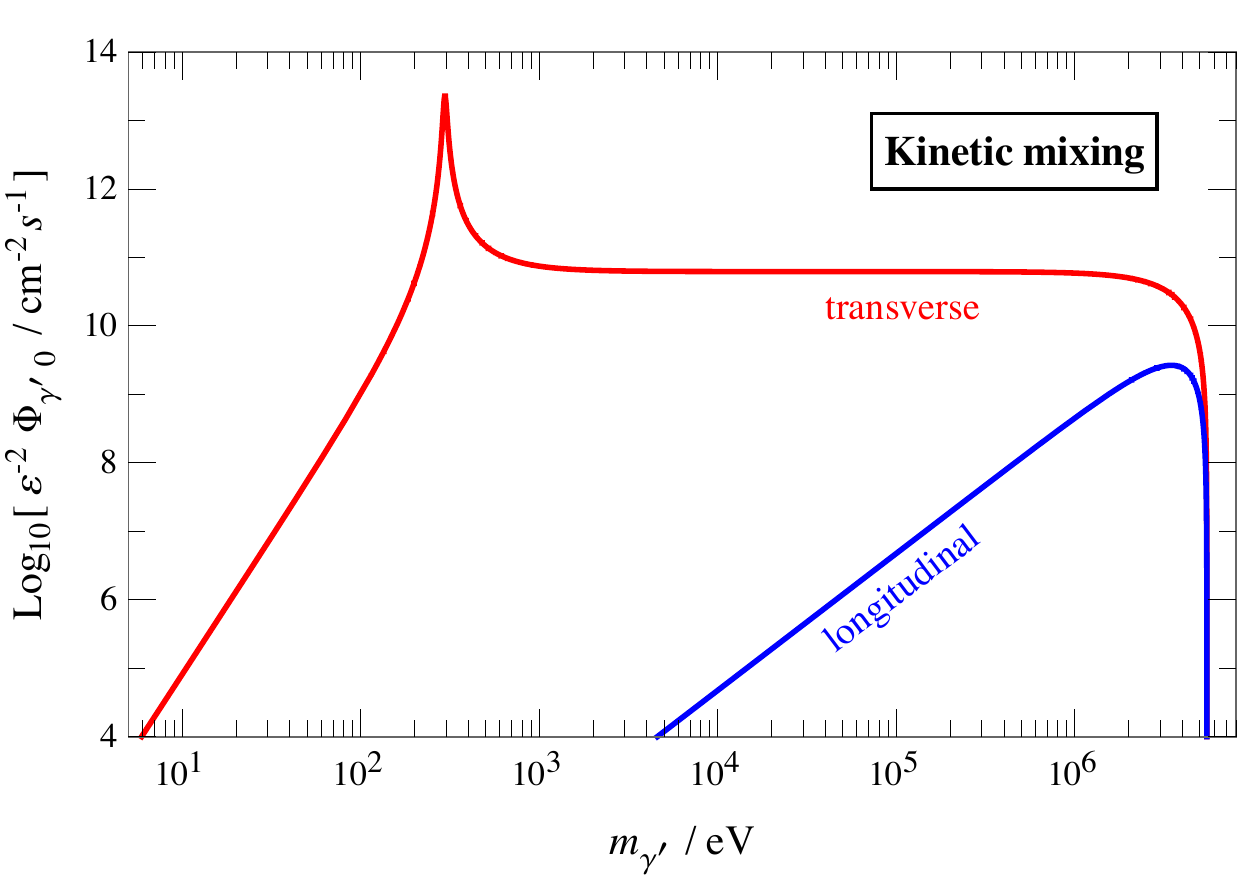}
\caption{\footnotesize Hidden vectors flux from the second state of the $pp$ chain for the kinetic mixing case ($\varepsilon \neq 0$). The red and blue lines indicate the flux of the transverse and longitudinal polarizations, respectively.}
\label{fig:DPflux}
\end{figure}

The expected fluxes $[\Phi_{\gamma^\prime 0}]_{\rm T,L}^{\varepsilon}$ of dark photons as a function of $m_{\gamma^\prime}$ and normalized in units of $\varepsilon^{2}$ are shown in Fig.~\ref{fig:DPflux} for transverse (red curve) and longitudinal (blue curve) polarizations. The former is provided by Eq.~\eqref{eq:DPFluxT} and the mass dependence enters only via the ratio $m_{\gamma^\prime}/\omega_{\rm pl}$. In the large mass region, $m_{\gamma^\prime} \gg \omega_{\rm pl}$, the production rate is independent of the mass. As we approach mass values $m_{\gamma^\prime} \simeq \omega_{\rm pl}$ the production is resonantly enhanced as manifest from the peak in the red curve in Fig.~\ref{fig:DPflux}. The width of such a resonance is determined by the imaginary part of $\pi_{\rm T}$ in Eq.~\eqref{eq:ImPiT}. Plasma screening effects suppress the production of transverse dark photons for $m_{\gamma^\prime} \ll  \omega_{\rm pl}$ via the power law behavior $(m_{\gamma^\prime} / \omega_{\rm pl})^4$. The discussion about the flux of longitudinal dark photons provided by Eq.~\eqref{eq:DPFluxL} is much simpler. First, there is no resonance peak for the reasons discussed in the previous section. Being away from the resonance implies also that the leading mass dependence is via the power law $(m_{\gamma^\prime} / \omega)^2$ which is the last factor in Eq.~\eqref{eq:DPFluxL}.

\subsubsection*{Production via plasma mixing}

The second case we investigate in our work is for a non-vanishing $U(1)^\prime$ charge of the electron (i.e., $q_e^\prime \neq 0$). As discussed in the previous section, this induces the in-medium mixing of dark gauge bosons with photons through a thermal electron loop. This effect is quantified by Eq.~\eqref{eq:Piggp}, and it provides an effective $\gamma^\prime$ couplings to the EM current~\cite{Hardy:2016kme}. Hence, even if the kinetic mixing is absent or sub-dominant, the $p(d,^3{\rm He})\gamma^\prime$ reaction could proceed via the plasma mixing. In this scenario, the flux of $5.49\,{\rm MeV}$ vectors also follows the EM transition rate and their explicit expressions can be obtained from the ones in Eqs.~\eqref{eq:DPFluxT} and \eqref{eq:DPFluxL} via the replacements $\varepsilon\,m_{\gamma^\prime}^2\rightarrow \frac{e^\prime q_e^\prime}{e q_e}\,\pi_{\rm T,L}$ inside the absolute values
\begin{align}
\label{eq:DPplasmaFluxT} \left[\Phi_{\gamma^\prime 0}\right]_{\rm T}^{q_e^\prime} \simeq & \, 
\frac{\phi_\gamma}{4\pi d_{\odot}^2} \times \frac{k^3}{\omega^3} \times 
\left|\frac{e^\prime q_e^\prime}{e q_e}   \frac{\pi_{\rm T}}{m_{\gamma^\prime}^2-\pi_{\rm T}}\right|^2 \ , \\
\label{eq:DPplasmaFluxL} \left[\Phi_{\gamma^\prime 0}\right]_{\rm L}^{q_e^\prime} \simeq & \,
\frac{\phi_\gamma}{4\pi d_{\odot}^2} \times \frac{k^3}{\omega^3} \times \left|\frac{e^\prime q_e^\prime}{e q_e}   \frac{\pi_{\rm L}}{m_{\gamma^\prime}^2-\pi_{\rm L}}\right|^2 \times \frac{1-P_{\rm M1}}{2} \times \frac{m_{\gamma^\prime}^2}{\omega^2} \ .
\end{align}

\begin{figure}[t!]
\centering
\includegraphics[width=0.7\textwidth]{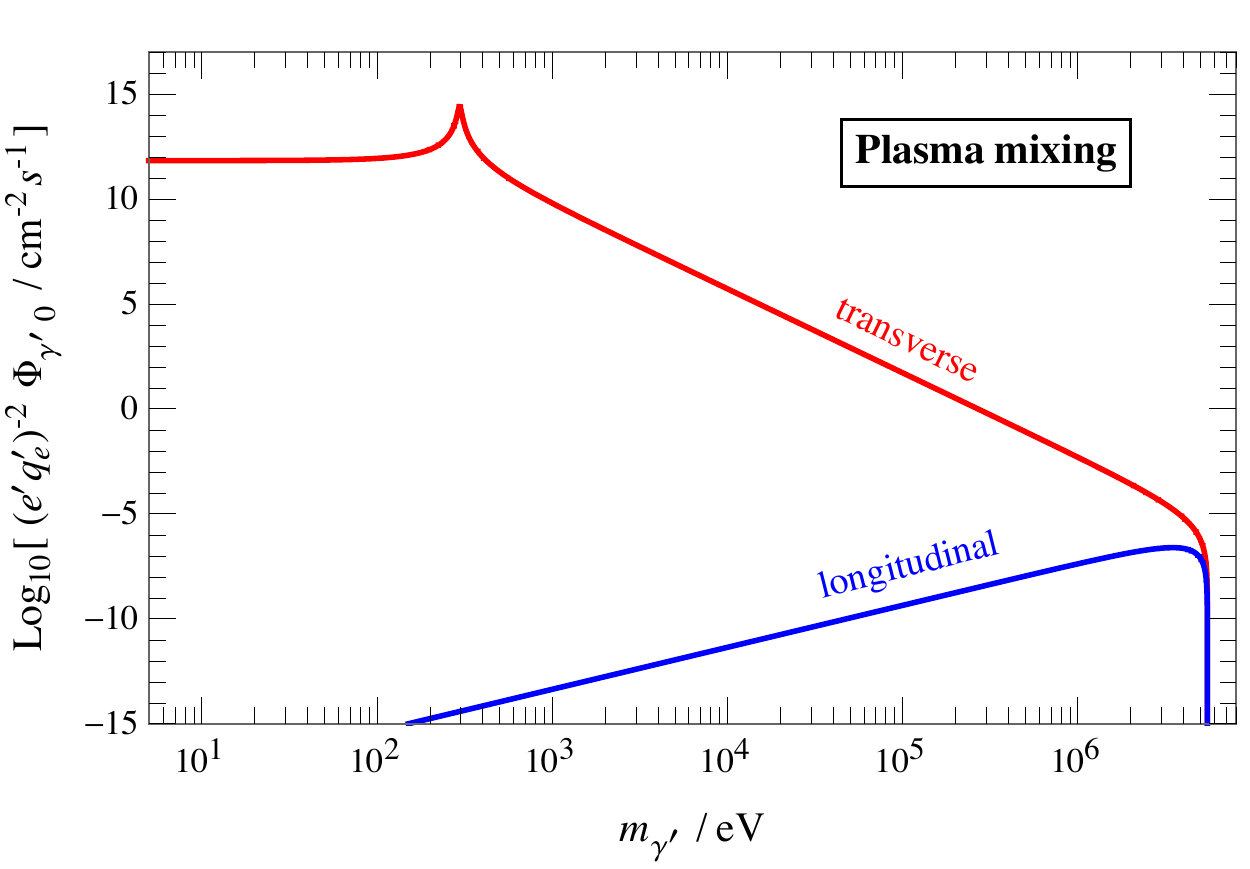}
\caption{\footnotesize Hidden vectors flux from the second state of the $pp$ chain for the plasma mixing case ($q^\prime_e \neq 0$). The red and blue lines indicate the flux of the transverse and longitudinal polarizations, respectively.}
\label{fig:BLflux}
\end{figure}

The expected fluxes for this case are shown in Fig.~\ref{fig:BLflux}. For transverse polarizations (red line), the flux is constant in the low mass region ($m_{\gamma^\prime}\ll \omega_{\rm pl}$), and it does not change until we reach $m_{\gamma^\prime} \simeq \omega_{\rm pl}$ where we find a resonant peak. At larger masses, it decreases according to the power law behavior $(\omega_{\rm pl} / m_{\gamma^\prime})^4$. The flux of longitudinal vectors (blue line) is proportional to the factor $(m_{\gamma^\prime}/ \omega)^2$, as it was in the dark photon scenario, but it is further suppressed by the factor $(\omega_{\rm pl}/\omega)^4$ due to plasma effects. 

\subsection{Attenuation of the incoming flux: solar absorption and decays}

The solar medium could potentially alter the propagation of $\gamma^\prime$ after production. As an example, the massive vectors can be reabsorbed into the medium near their production point via Compton-like scatterings. The absorption rate can be estimated by combining the imaginary part of the transverse photon self-energy in Eq.~\eqref{eq:ImPiT} with the medium effective $\gamma^\prime$ coupling to the electron in Eq.~\eqref{eq:DPeffMed}, and it approximately results in $(\varepsilon e -e^\prime q_e^\prime)^2 |m_{\gamma^\prime}^2/(m_{\gamma^\prime}^2-\pi_{\rm T,L})|^2 (\alpha n_e/3 m_e^2)$. In the standard solar model~\cite{Bahcall:1987jc,Bahcall:1989ks}, the electron number density decreases exponentially $n_e(r)\propto\exp [-10.54 \,r/R_\odot ]$ along the radial direction~\cite{Bahcall:2000nu}, where $R_\odot = 6.96 \times 10^{10}\,{\rm cm}$ is the solar radius. Thus, the absorption mostly occurs near the innermost region of the Sun. The optical length is written by
\beq
\begin{split}
\tau_{\gamma^\prime} & \sim 
\, \int_0^{R_\odot} dr\,  \left(\varepsilon e - e^\prime q_e^\prime\right)^2\left|\frac{m_{\gamma^\prime}^2}{m_{\gamma^\prime}^2-\pi_{\rm T,L}}\right|^2\frac{\alpha}{3m_e^2}n_e\left(r\right) = \\ & 
= \mathcal{O}(0.1) \, \left(\varepsilon e - e^\prime q_e^\prime\right)^2\left|\frac{m_{\gamma^\prime}^2}{m_{\gamma^\prime}^2-\pi_{\rm T,L}}\right|^2_{r=0}\frac{\alpha}{3m_e^2}n_e\left(0\right) R_\odot \ .
\end{split}
\eeq
Moreover, transverse vectors could resonantly oscillate with photons when the resonance condition $m_{\gamma^\prime} = \omega_{\rm pl}$ is achieved, while no resonance effect is present for the longitudinal polarization as discussed around Eq.~\eqref{eq:ImPiT}. Such a resonant conversion probability relies on the adiabaticity at the resonance point, which is parametrized by~\cite{Mirizzi:2009iz,Choi:2019jwx}
\bea
\gamma_{\rm ad} = 2 \left(\varepsilon e - e^\prime q_e^\prime\right)^2 \left|\frac{d \log \omega_{\rm pl}^2}{d r}\right|^{-1} \frac{m_{\gamma^\prime}^2}{\omega} = \mathcal{O}\left(0.1\right)\, \left(\varepsilon e - e^\prime q_e^\prime\right)^2  \frac{m_{\gamma^\prime}^2}{\omega} R_\odot \ ,
\eea
where we use $\omega_{\rm pl}^2\propto n_e$ in the second relation. The conversion probability reads as
\bea
P_{\gamma^\prime \leftrightarrow \gamma}^{\rm res} \simeq 1- \exp\left(-\frac{\pi \gamma_{\rm ad}}{2}\right) \, ,
\eea
where the second term corresponds to the level-crossing probability described by the Landau-Zener formula~\cite{Landau:1932ab,Zener:1932ws,Parke:1986jy}. Due to the exponential change of $\omega_{\rm pl}^2$ along the radial direction, a resonant conversion becomes non-adiabatic (i.e., $\gamma_{\rm ad} \ll 1$) at $|\varepsilon e - e^\prime q_e^\prime| < \mathcal{O}(10^{-6})$.

Even if the hidden vectors manage to escape the Sun, they can still decay before reaching the Earth. The net result is to diminish effectively their incoming flux whether the decays are visible or not. A decay channel that is always kinematically allowed is the one into three photons, $\gamma^\prime \rightarrow \gamma\gamma\gamma$, and the process proceeds via an electron loop~\cite{Pospelov:2008jk}. At masses above the kinematical threshold $(m_{\gamma^\prime} > 2m_e)$, massive vectors can decay into electron-positron pairs again through the coupling to electron. If couplings to neutrino are present (e.g., gauging lepton numbers), there would be additional decay channels into  neutrino-antineutrino pairs, $\gamma^\prime \rightarrow \nu_i \bar{\nu}_i$. Laboratory searches ~\cite{KATRIN:2019yun,KATRIN:2021uub} and cosmological analysis~\cite{Planck:2018vyg} bound the neutrino mass to $m_{\nu} < \mathcal{O}(1)\,{\rm eV}$, and massive vectors in the broad range $2 m_\nu \leq m_{\gamma^\prime} \leq 2 m_e$ can decay to this additional channel.  We find it useful to identify the electron coupling because it plays a crucial role for decays as well as for the detection discussed in the next subsection. The low density in both cases allows us to neglect collective effects and use the Lagrangian for physical states in the vacuum in Eq.~\eqref{eq:DPeffVac}. Focusing only on the electron coupling, we find
\bea
\left. \mathcal{L} \right|^{\rm (electron)}_{\rm vacuum} = g_e \,  \bar{e} \gamma^\mu e A^\prime_\mu \ , \qquad \qquad \qquad g_e = -\varepsilon e + e^\prime q_e^\prime  \ .
\label{eq:Ldetec}
\eea
The decay width for the three-photon final state explicitly reads as
\bea
\Gamma_{\gamma^\prime \rightarrow \gamma\gamma\gamma} \approx (4.7 \times 10^{-8}) 
\frac{g_e^2 \,\alpha^3}{4\pi} \frac{m_{\gamma^\prime}^9}{m_e^8}\ ,
\eea
and it is always negligible in our analysis. The width to electron-positron pairs evaluated in the vector rest frame, and averaged over the three polarizations, results in
\bea
\Gamma_{\gamma^\prime \rightarrow e^- e^+}  =  \frac{g_e^2 \, m_{\gamma^\prime}}{12\pi} \left(1+\frac{2m_e^2}{m_{\gamma^\prime}^2}\right)\sqrt{1-\frac{4m_e^2}{m_{\gamma^\prime}^2}} \, \Theta \left(m_{\gamma^\prime}-2m_e\right)\, ,
\label{eq:DPdecay}
\eea
where $\Theta(x)$ is the Heaviside step function. If couplings to neutrinos are present, the rate calculation is analogous to the one leading to Eq.~\eqref{eq:DPdecay} with replacement of $g_e \rightarrow g_\nu$ and $m_e \rightarrow m_{\nu}$, where $g_\nu$ and $m_\nu$ are the $\gamma^\prime$ coupling to the neutrino and the neutrino mass, respectively, and an additional $1/2$ factor due to the chiral (left-handed) coupling.

Reabsorptions and conversions of $\gamma^\prime$ across the solar medium deplete the fluxes given in the previous section by an overall factor of $\exp[-(\tau_{\gamma^\prime}+ \pi \gamma_{\rm ad}/2)]$. In addition, the effective flux reaching the terrestrial detector gets another overall correction accounting for decays. The flux that eventually reaches terrestrial detectors results in
\bea
\left[ \Phi_{\gamma^\prime} \right] = \, \exp\left[- \left(\tau_{\gamma^\prime} + \frac{\pi}{2} \gamma_{\rm ad} \right)\right]\, \exp\left[- \Gamma_{\gamma^\prime} \frac{m_{\gamma^\prime}}{\omega} \frac{d_\odot}{v_{\gamma^\prime}}\right] \, \left[ \Phi_{\gamma^\prime 0}\right] \ .
\label{eq:phicorrected}
\eea
Here, $\left[\Phi_{\gamma^\prime 0}\right]$ was derived in Sec.~\ref{sec:production} for the different cases, $\Gamma_{\gamma^\prime}$ denotes the total $\gamma^\prime$ decay width including both visible and invisible channels, the ratio $m_{\gamma^\prime}/\omega$ is a Lorentz time dilatation factor, and $v_{\gamma^\prime} = \sqrt{1-m_{\gamma^\prime}^2/\omega^2}$ is the vector velocity. Resonant conversions dominate the opacity in the mass range $10^{-2} \leq m_{\gamma^\prime}/\omega_{\rm pl} \leq 1$, reabsorption via Compton-like scatterings controls the redistribution of $\gamma^\prime$'s energy into the medium otherwise. 

\subsection{Terrestrial detection}
\label{sec:detection}

\begin{figure}[t!]
\centering
\includegraphics[width=0.7\textwidth]{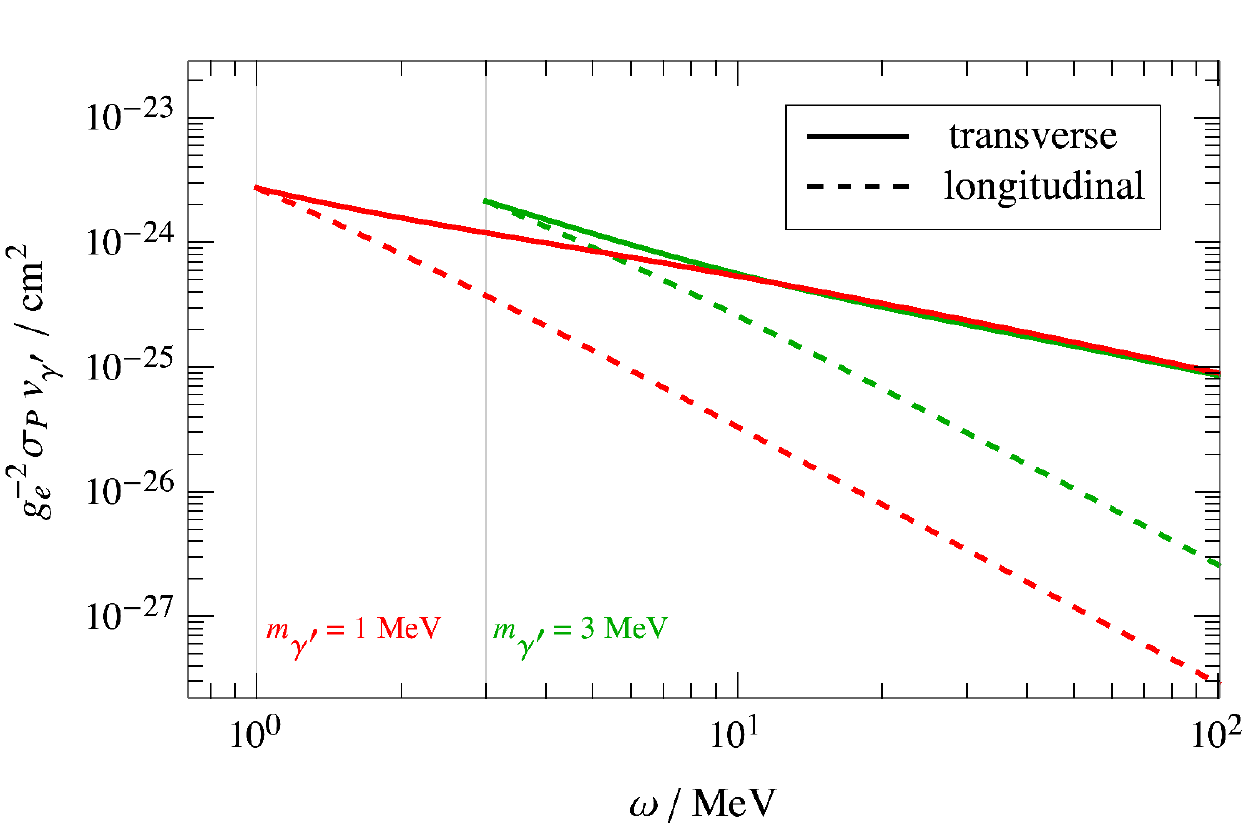}
\caption{\footnotesize Cross section $\sigma_P$ of the Compton-like scattering process $\gamma^\prime \, + \, e \, \rightarrow \, \gamma \, + \, e$, multiplied by the $\gamma^\prime$ velocity $v_{\gamma^\prime}$, as a function of the incident $\gamma^\prime$ energy $\omega$ for $m_{\gamma^\prime} = 1\,{\rm MeV}$ (red) and $3\,{\rm MeV}$ (green). The solid and dashed lines correspond to transverse ($P = {\rm T}$) and longitudinal ($P = {\rm L}$) polarizations, respectively.}
\label{fig:DPCross}
\end{figure}

Once the massive vectors reach finally terrestrial laboratories, there are in principle several  processes allowing for their detection. The leading one is the Compton-like scattering  
\bea
\gamma^\prime \, + \, e \, \rightarrow \, \gamma \, + \, e \, .
\eea
The detection rate is controlled by the combination $g_e = (-\varepsilon e + e^\prime q_e^\prime)$ identified in Eq.~\eqref{eq:Ldetec}. We derive the cross sections $\sigma_P$ for transverse ($P={\rm T}$) and longitudinal ($P={\rm L}$) polarizations in Appendix~\ref{app:Compton}, providing their expressions in Eqs.~\eqref{eq:sigmaT} and \eqref{eq:sigmaL}, respectively. The final expressions are rather long and we do not report them here. Once the hidden vector mass becomes negligible (i.e., $m_{\gamma^\prime} \ll \omega, m_e$), the expressions simplify 
\begin{align}
\sigma_T \simeq & \,  8.5 \, g_e^2 \, \times 10^{-25} \, {\rm cm}^2 \ , \\
\sigma_L \simeq & \, 1.2 \, g_e^2 \, \left(\frac{m_{\gamma^\prime}}{1 \, {\rm keV}}\right)^2 \, \times 10^{-31} \, {\rm cm}^2 \ .
\end{align}
We take the massless limit for transverse polarizations, whereas for the longitudinal ones we keep only the leading term in the small vector mass. The initial state electron is at rest in the laboratory frame, and the (square root of the) energy in the center of mass quantified by the Mandelstam variable $s$ reads $s = m_e^2 + m_{\gamma^\prime}^2 + 2m_e \omega$. We show in Fig.~\ref{fig:DPCross} the product between the vector initial velocity $v_{\gamma^\prime}$ and the cross section $\sigma_P$ of the Compton-like scattering process for transverse (solid) and longitudinal (dashed) polarizations. We fix the vector mass to two different values: $m_{\gamma^\prime} = 1\,{\rm MeV}$ (red) and $m_{\gamma^\prime} = 3\,{\rm MeV}$ (green). The scattering rate of all polarizations is equivalent when $\omega = m_\gamma^\prime$; in such a limit the vector is at rest in the rest frame of the target electron, and therefore the transverse and longitudinal polarizations are indistinguishable due to the isotropic environment of the detector.

Besides mediating Compton-like scattering, the coupling to the electron in Eq.~\eqref{eq:Ldetec} could also induce decays to $e^\pm$ pairs once they become kinematically allowed. These visible decays during their propagation inside the detector would provide an additional detection channel. However, this is never important for our analysis because once visible decays become available the net flux detected on Earth is drastically reduced. Apart from electrons, hidden vectors can also scatter coherently with ionized nuclei (e.g., carbon), but their rate is negligible due to the suppression by heavy nuclear masses. In principle, massive vectors can be absorbed also via other scattering processes. For instance, in a process similar to the photoelectric effect $\gamma + e \rightarrow e^*$, the vector is absorbed and an electron $e^*$ is emitted with an energy equal to the difference between the absorbed boson energy and the electron binding energy. The cross section for this process is approximately $\sigma_{\gamma^\prime e \rightarrow e^*} \approx \sigma_{\gamma e \rightarrow e^*} g_e^2 \, ( 3\omega^2/16\pi\alpha m_e^2) \, v_{\gamma^\prime}^{-1}(1-v_{\gamma^\prime}^{2/3}/3)$ with $\sigma_{\gamma e \rightarrow e^*}$ the cross section for the photoelectric effect at the energy $\omega$. Hidden vectors can produce electron-positron pairs in the electric field of nuclei or electrons but these processes are subdominant with respect to the Compton scattering in the JUNO detector (see Fig.~1 in Ref.~\cite{Lucente:2022esm}).

\section{Hidden vectors at JUNO}
\label{sec:JUNOdet}

The primary goal of the next-generation multi-purpose underground liquid scintillator (LS) JUNO detector is to resolve the neutrino mass ordering. Because of its excellent energy resolution capability and the large fiducial volume (FV), the JUNO detector can be used also to test various new physics predictions, such as proton decay, neutrino non-standard interactions, and hidden sector particles. In this section, we investigate the JUNO sensitivity to the new physics scenarios presented in this work.

The main features of JUNO have been thoroughly described in Refs.~\cite{JUNO:2015zny,JUNO:2022lpc}. The LS detector has a fiducial mass of 20 kton of Linear
alkyl benzene (LAB), C$_{19}$H$_{32}$, doped with $3$ g/L of $2,5$-diphenyloxazole (PPO) and $15$ mg/L of p-bis-(o-methylstyryl)-benzene (bis-MSB), contained in a sphere of radius $17.7$ m. The energy resolution is $\bar{\sigma}/E= 3\%/\sqrt{E}$, where both $\bar{\sigma}$ and $E$ are in MeV.
To detect $^8$B solar neutrinos with threshold energy of $2$ MeV, an energy-dependent FV cut is considered (see Ref.~\cite{JUNO:2020hqc} for more details). In particular, for events with energy larger than 5 MeV, a FV of 16.2 kton contained in a sphere of radius $r = 16.5$ m is assumed. After all the selection cuts described in~\cite{JUNO:2020hqc}, about 60,000 solar neutrino events and 30,000 radioactive background events are expected in
10 years of data taking, as shown in Table 4 and Fig. 11 of Ref.~\cite{JUNO:2020hqc}. In our work, both types of events represent the background of the solar dark gauge boson signal.

The detection rate at JUNO can be quantified according to what detection channel is dominant. For our study, Sec.~\ref{sec:detection} discusses how the detection proceeds via Compton-like scattering only, and the expected number of events per time interval $\Delta t$ results in
\begin{equation}
  \left.\frac{\Delta{N}_{\rm ev}}{\Delta t}\right|_{\rm Compton-like} =  \left[\Phi_{\gamma^\prime}\right] \otimes N_{\rm tar} \otimes \sigma \otimes \mathcal{R} \otimes \epsilon\,,
\label{eq:nevdt}    
\end{equation}
where $\left[\Phi_{\gamma^\prime}\right]$ is the net vector flux on Earth given by Eq.~\eqref{eq:phicorrected}, $N_{\rm tar}$ the number of targets, and $\sigma$ the scattering cross section. Here, $\mathcal{R}$ and $\epsilon$ represent the detector energy resolution and the detector efficiency, respectively, and we assume $\epsilon = 1$ for all detection channels over the energy threshold throughout this work. We report for completeness also the decay rate if visible decays (such as $3\gamma$ or $e^\pm$) are allowed inside the detector 
\begin{equation}
  \left.\frac{\Delta{N}_{\rm ev}}{\Delta t}\right|_{\rm Decays}  = \left[\Phi_{\gamma^\prime}\right] \otimes \frac{V_f}{l_i} \otimes \epsilon\,,
    \label{eq:nevdtdec}
\end{equation}
where $V_f$ is the detector FV and $l_i$ is the decay length in the laboratory frame for $i-$th decay channel. Combining everything together, the total number of expected events reads 
\bea
N_{\rm ev}  =  \sum_{P={\rm T,L}}\left[\Phi_{\gamma^\prime}\right]_P \,\left[ N_e \sigma_P + 
\frac{V_f}{v_{\gamma^\prime}}\, \frac{m_{\gamma^\prime}}{\omega} \, 
\Gamma_{\gamma^\prime}{\textrm{Br}}\left(\gamma^\prime \rightarrow {\rm vis}\right) \right] \,  t  \ .
\label{eq:nev}
\eea
Here, $N_{e}\simeq 5.5 \times 10^{33}$ is the total number of electrons in the $16.2$ kton FV ($V_f$) of the spherical structure with the radius $16.5\,{\rm m}$, implying $V_f = 1.9 \times 10^4 \,{\rm m}^3$, and $t = 10$~years is the measurement time duration. We derive the Compton-like scattering cross section in Appendix~\ref{app:Compton}. The total decay rate $\Gamma_{\gamma^\prime}$, which is multiplied by the factor $(m_{\gamma^\prime} / \omega)$ to account for the Lorentz dilation, includes both visible and invisible channels, and $v_{\gamma^\prime}$ denotes the vector velocity in the laboratory frame. The branching ratio to visible decay channels ${\textrm{Br}}(\gamma^\prime \rightarrow {\rm vis})$ in Eq.~\eqref{eq:nev} is required since only visible decays lead to detectable events. 

\subsection{Fitting procedure}
\label{sec:Numerical}

We evaluate now the future reaches of JUNO detector on the hidden vector models. We adopt the likelihood prescription developed in \cite{Lucente:2022esm} for our numerical analysis, and we briefly summarize below the main steps. In order to forecast the detector sensitivity, we express the $\chi^2$ function which contains the information on dark photons as
\begin{equation}
\begin{split}
\chi^2= & \, 2\times\sum_{i}\left(  N_{i,\rm pre}-N_{i,\rm exp}+N_{i,\rm exp}\times \log\frac{N_{i,\rm exp}}{N_{i,\rm pre}}\right)  +\left( \frac{\epsilon_{\rm sb}}{\sigma_{\rm sb}}\right) ^2+\left( \frac{\epsilon_{\rm rb}}{\sigma_{\rm rb}}\right) ^2, \\
 N_{i,\rm pre} &= (1+\epsilon_{\rm sb})\times B_{i,\rm sb} +  (1+\epsilon_{\rm rb})\times B_{i,\rm rb} + \dfrac{S}{\sqrt{2\pi}\bar{\sigma}}\times e^{-\frac{(\bar{E}-E_i)^2}{2\bar{\sigma}^2}},
  \end{split}
\label{equ:spectrumT}
\end{equation}
where $N_{i,\rm exp}$ and $N_{i,\rm pre}$  represent the expected solar neutrino events and the predicted number of events in the presence of dark photon corresponding to the $i^{th}$ energy bin, with energy $E_i$~\cite{JUNO:2020hqc},  respectively. Also, $B_{i,\rm sb}$ and $B_{i,\rm rb}$ represent the solar neutrino and the radioactive background events,\footnote{We adopt $N_{i,{\rm exp}} = B_{i,{\rm sb}} + B_{i,{\rm rb}}$.} while $ \epsilon_{sb} $ and $ \epsilon_{rb} $ are the nuisance parameters, and the corresponding solar and radioactive background normalization uncertainties are given by $ \sigma_{\rm sb} = 5\%  $ and $ \sigma_{\rm rb} = 15\% $, respectively. We model the new physics contribution by using a Gaussian function,  with $S$ parameterizing the expected hidden vector peak intensity centered at $\bar{E}=5.49$~MeV and with a width $\bar{\sigma} = 0.07$~MeV (see Fig. 2 of \cite{Lucente:2022esm} for details). Through a $ \chi^{2} $ test, we find that the projected JUNO sensitivity at the 90\% confidence level (C.L.) is $S_{\rm lim} = 97$ counts in 10 years, as shown by Fig. 3 in \cite{Lucente:2022esm}.

\subsection{Results}
\begin{figure}[t!]
\centering
\includegraphics[width=0.65\textwidth]{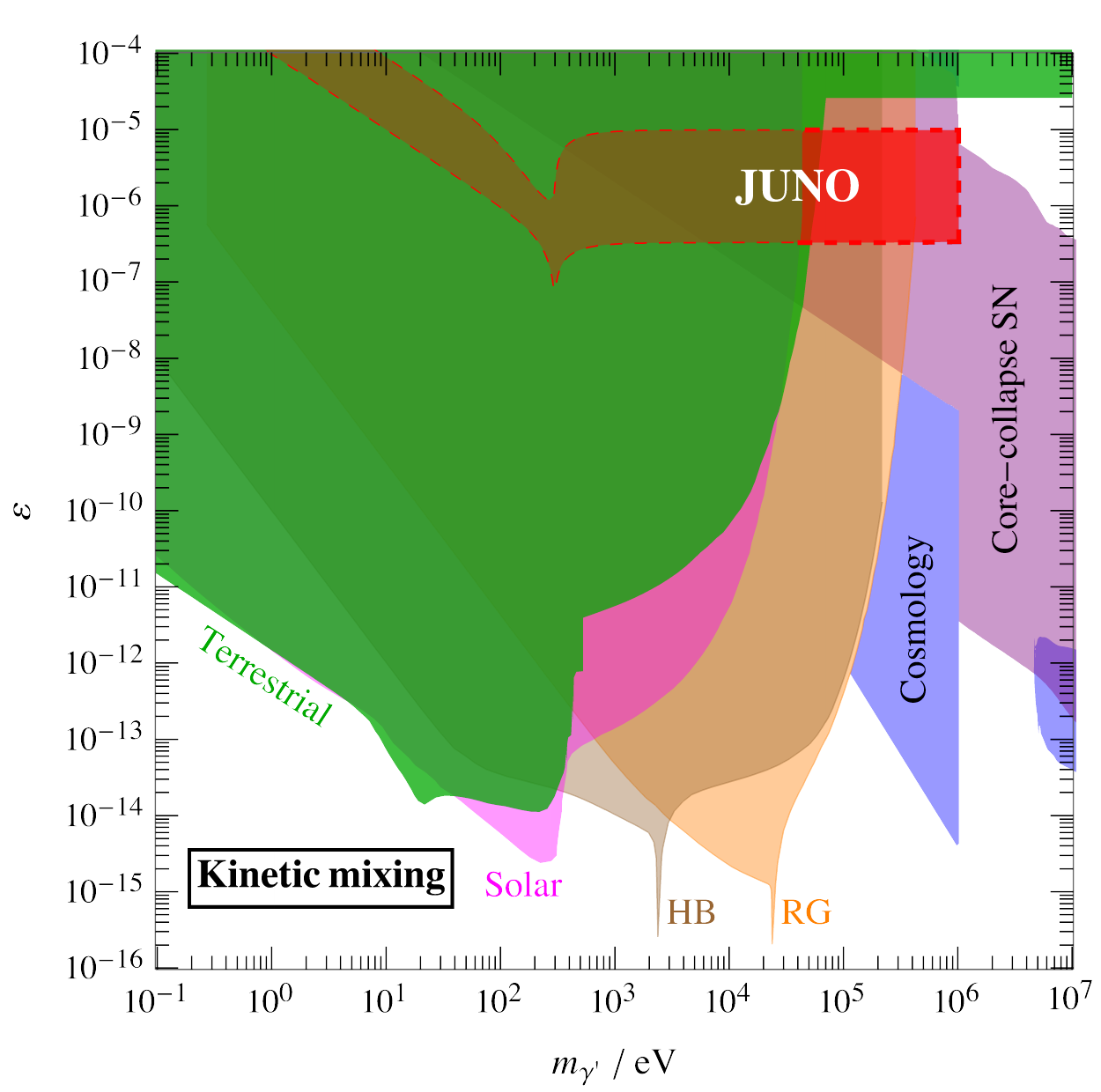}
\caption{\footnotesize Exclusion regions and projected JUNO sensitivity for the dark photon scenario communicating with the SM via a non-vanishing kinetic mixing ($\varepsilon \neq 0$). In the $(m_{\gamma^\prime}, \varepsilon)$ plane, the red region with a dashed boundary represents the discovery reach of the JUNO experiment at the 90\% C.L. sensitivity. The blue and green regions indicate the current constraints based on cosmology and terrestrial experiments, respectively. The magenta, brown, orange, and purple regions are excluded by astrophysics with respect to the Sun, horizontal branch stars~\cite{Li:2023vpv}, red giants~\cite{Dolan:2023cjs}, and core-collapse supernovae~\cite{Chang:2016ntp,DeRocco:2019njg,Hong:2020bxo,Calore:2021lih,Caputo:2022mah,Shin:2022ulh}, respectively; see Ref.~\cite{Caputo:2021eaa} and the references therein for more details.}
\label{Fig:JUNO_DP}
\end{figure}
%
We present the JUNO discovery reach by starting with the dark photon scenario. For this case, the only communication channel with the visible sector is a non-vanishing kinetic mixing parameter ($\varepsilon \neq 0$). Fig.~\ref{Fig:JUNO_DP} shows an overview of the parameter space. We show in red the region that will be probed by the JUNO at 90\% C.L. sensitivity. For masses $m_\gamma^\prime > 2\,m_e$ dark photons can decay into electron-positron pairs, and the corresponding attenuation depletes exponentially the vector flux for values of the kinetic mixing parameter that in principle would allow for detection. Thus, JUNO cannot probe dark photon masses above the MeV scale, explaining the sharp cut at $m_a \approx 1$~MeV in the JUNO sensitivity region in Fig.~\ref{Fig:JUNO_DP}. On the contrary, for masses below the electron-positron threshold, JUNO will probe kinetic mixing parameters $3 \times 10^{-7} \lesssim \varepsilon \lesssim 10^{-5}$ for masses in the range $\omega_{\rm pl} \simeq 300~{\rm eV} \lesssim m_{\gamma^\prime} \lesssim 2 m_e \simeq 1\,{\rm MeV}$. For $m_{\gamma^{\prime}}\lesssim \omega_{\rm pl}$ the lower limit of the JUNO sensitivity scales linearly with the mass since $[N_{ev}]^{\varepsilon}\propto \varepsilon^4 m_{\gamma^\prime}^4$ (see Eq.~\eqref{eq:DPFluxT}). The deep at $m_{\gamma^\prime}\approx \omega_{\rm pl}$ is due to the resonance at $m_{\gamma^\prime} = \omega_{\rm pl}$. Finally, for $m_{\gamma^\prime}\gtrsim \omega_{\rm pl}$ the event rate is independent of the dark photon mass since $[N_{ev}]^{\varepsilon} \propto \varepsilon^4$. JUNO will provide the best terrestrial bounds for the dark photon scenario in the mass range $ 50~{\rm keV} \lesssim m_{\gamma^\prime} \lesssim 1\,{\rm MeV}$. 

As shown in Fig.~\ref{Fig:JUNO_DP}, JUNO will be sensitive to a region already excluded by astrophysical arguments. In particular, for $m_{\gamma^\prime} \gtrsim 50$~keV, where JUNO will provide the strongest terrestrial bound, the projected limit overlaps with the bounds from Horizontal Branch (HB) stars, Red Giants (RG) and core-collapse supernovae (SNe). The HB and the RG constraints have been recently revised in Ref.~\cite{Dolan:2023cjs}, using simulations with a self-consistent inclusion of dark photons as an additional energy-loss channel, focusing on the region of the parameter space with $\varepsilon < 10^{-13}$ where dark photons freely escape the stellar interior. For the large values of the kinetic mixing which can be probed by JUNO ($\varepsilon \gtrsim 10^{-7}$), the hidden vector reabsorption in the stellar interior cannot be neglected. Since stellar-evolution simulations considering dark-photon reabsorption are still missing, for $\varepsilon > 10^{-13}$ the HB and RG bounds shown in Fig.~\ref{Fig:JUNO_DP} are obtained extrapolating the ones from Ref.~\cite{Dolan:2023cjs}. 
On the other hand, the SNe constraints~\cite{Chang:2016ntp,DeRocco:2019njg,Hong:2020bxo,Calore:2021lih,Caputo:2022mah,Shin:2022ulh} shown in Fig.~\ref{Fig:JUNO_DP} take into account the trapping regime, where produced hidden vectors do not freely escape the SN core. Nonetheless, such analyses are subject to large theoretical and observational uncertainties (see, e.g., Ref.~\cite{Fiorillo:2023frv} for a recent discussion on the issues related to the observed events from SN 1987A) and no SN simulation with a self-consistent inclusion of trapped hidden vectors is available. As a result, it is worthwhile to probe this region of the parameter space with different approaches.

We note that the $5.49\,{\rm MeV}$ solar hidden vector flux is evaluated assuming the standard solar model without changes due to extra hidden vector emission. 
This may lead to inconsistency with the evolutionary history of the Sun~\cite{Vinyoles:2015aba,Li:2023vpv}.
A significant hidden vector production inside the Sun can affect the solar evolution and leave imprints on some observables, such as the $^8{\rm B}$ and $^7{\rm Be}$ neutrino fluxes and the helioseismologic quantities.
The magenta region in Fig.~\ref{Fig:JUNO_DP} corresponds to such constraints based on the solar evolution, which rely on the solar hidden vector luminosity dominantly distributed in the range of the core temperature ($\sim{\rm keV}$). In this context, the $5.49\,{\rm MeV}$ hidden vector spectrum contributes to the luminosity as $\phi_{\gamma} (k^3/\omega^2) \varepsilon^2 |m_{\gamma^\prime}^2/(m_{\gamma^\prime}^2-\pi_{\rm T})|^2 \sim \varepsilon^2 |m_{\gamma^\prime}^2/(m_{\gamma^\prime}^2-\pi_{\rm T})|^2 L_\odot$. Therefore, the evaluation of the JUNO sensitivity at masses above $10\,{\rm keV}$, corresponding to the brighter red-colored region, is self-consistent in terms of the solar profile.

\begin{figure}[t!]
\centering
\includegraphics[width=0.65\textwidth]{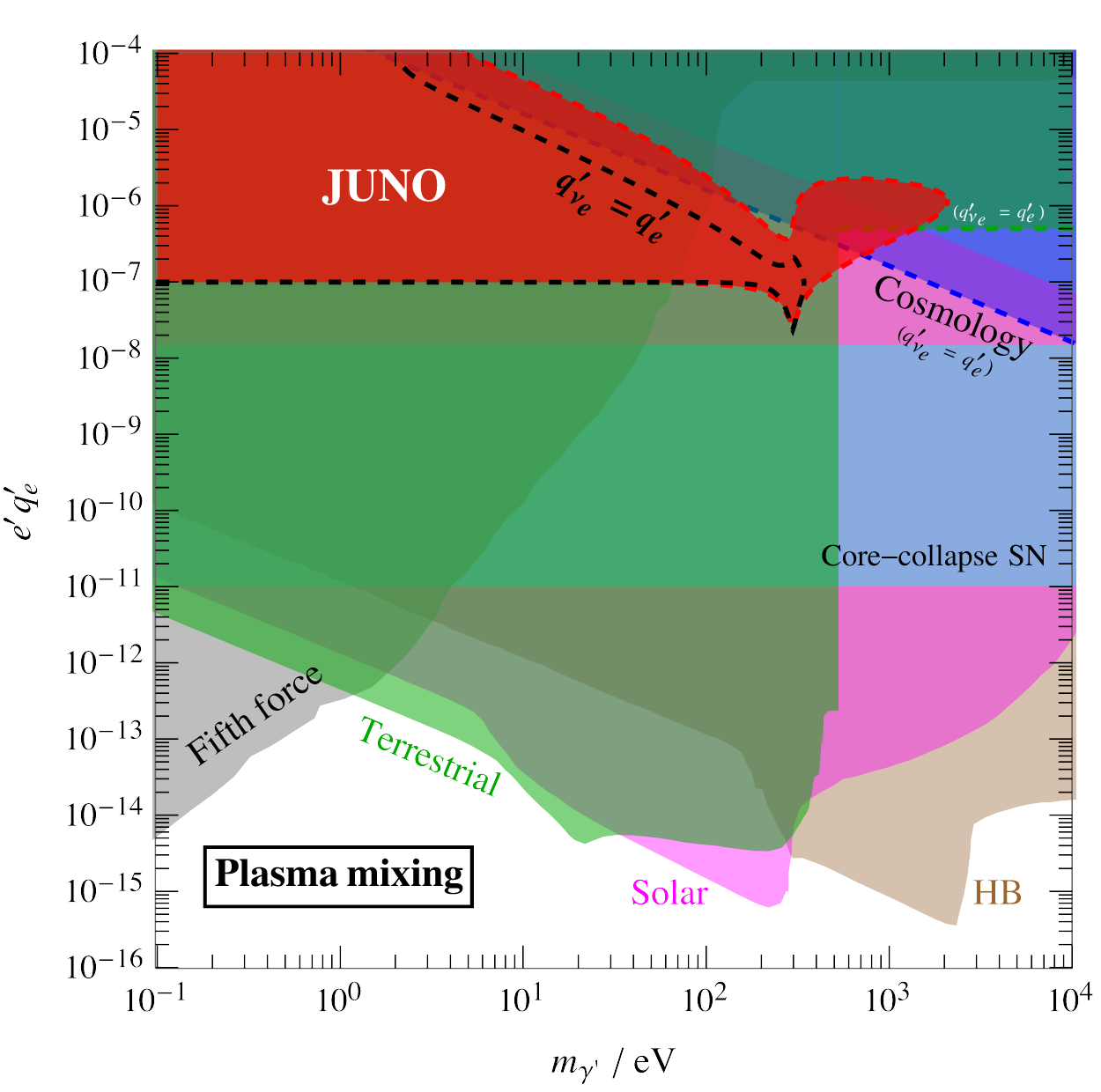}
\caption{\footnotesize Exclusion regions and projected JUNO sensitivity for the dark gauge boson scenario communicating via plasma mixing ($q_e^\prime \neq 0$). For this case, we have two different dashed boundaries for the red regions representing the discovery reach of the JUNO experiment at the 90\% C.L. sensitivity. The dashed red and dashed black boundaries correspond to the case of $q_{\nu_e}^\prime = 0$ and $q_{\nu_e}^\prime = q_e^\prime$, respectively. The gray and green regions indicate the current constraints based on fifth force tests and terrestrial experiments, respectively. The blue and dashed green regions come from the cosmological bounds and reactor experiments for $q_{\nu_e}^\prime = q_e^\prime$; see the main text for more details. The magenta, brown, and cyan regions are excluded by astrophysics with respect to the Sun, horizontal branch stars, and core-collapse supernovae, respectively.}
\label{Fig:JUNO_IM}
\end{figure}

Fig.~\ref{Fig:JUNO_IM} reports the exclusion plot for the plasma mixing scenario ($q_e^\prime \neq 0$). In this case, a non-vanishing coupling $q_\nu^\prime$ of the massive vector to SM neutrinos can alter the projected sensitivity. The red region delimited by the red dashed line is the projected sensitivity at $90\%$ C.L. for $q_{\nu_e}^\prime = 0$, while the dashed black line delimits the sensitivity in the case of $q_{\nu_e}^\prime = q_e^\prime$ (in agreement with electroweak gauge invariance). For our analysis, we pick $m_{\nu} = 1\,{\rm eV}$ as a fiducial value. As in the kinetic mixing theory, the parametric behavior of the lower part of the JUNO sensitivity for the in-medium mixing theory depends on the flux $\Phi_{\gamma^\prime 0}$ given in Eq.~\eqref{eq:DPplasmaFluxT} with an additional $(e^{\prime}q_e^\prime)^2$ factor from detection. For $m_{\gamma^\prime}<\omega_{\rm pl}$ the JUNO sensitivity is independent of the dark gauge boson mass since $[N_{ev}]^{q_e^\prime}\propto (e^{\prime}q_e^\prime)^4$.

Also in this case, there is a deep at $m_{\gamma^\prime}\approx \omega_{\rm pl}$ and for larger masses the JUNO sensitivity scales as ${e^{\prime}q_e^\prime \propto m_{\gamma^\prime}}$ since $[N_{ev}]^{q_e^\prime}\propto (e^{\prime}q_e^\prime)^4 m_{\gamma^\prime}^{-4}$ for $m_{\gamma^\prime} > \omega_{\rm pl}$. Decays into neutrino pairs allowed for $q_{\nu_e}^\prime \neq 0$ suppress the net flux, thus the upper limit scales as $e^\prime q_e^\prime \propto m_{\gamma^\prime}^{-1}$. In the absence of coupling to neutrinos, hidden vectors could still be absorbed into the solar medium. In particular, for $m_{\gamma^\prime}< \omega_{\rm pl}$ they are mainly reabsorbed after resonant conversions, leading to an upper limit which scales as $e^\prime q_e^\prime \propto m_{\gamma^\prime}^{-1}$. On the other hand, for $m_{\gamma^\prime}> \omega_{\rm pl}$ the main absorption channel is the Compton-like scattering, which limits the JUNO sensitivity to values $e^   \prime\,q_e^\prime \lesssim 2\times 10^{-6}$. In Fig.~\ref{Fig:JUNO_IM}, the magenta and brown regions show astrophysical constraints based on the stellar evolution of the Sun and HB stars~\cite{Harnik:2012ni,Heeck:2014zfa,Hardy:2016kme}, respectively, and the cyan region comes from observables related to core-collapse SNe~\cite{DeRocco:2019njg,Shin:2022ulh} and neutron star cooling~\cite{Hong:2020bxo,Shin:2021bvz,Shin:2022ulh}. In addition, the gravity experiments testing fifth forces lead to the stringent constraints at masses below $\mathcal{O}(1)\,{\rm eV}$~\cite{Fischbach:1999bc,Smith:1999cr,Adelberger:2009zz,Wagner:2012ui,Murata:2014nra,Fayet:2017pdp}, depicted by the gray region. The current terrestrial bounds from XENON1T~\cite{An:2020bxd} exclude the green-colored region.
For coupling to neutrinos ($q_{\nu}^\prime \neq 0$), there would be also cosmological constraints from Big Bang nucleosynthesis~\cite{Heeck:2014zfa,Knapen:2017xzo} that would exclude the blue region delimited by the dashed blue line (to derive this constraint we set the electroweak preserving charge assignment $q_{\nu}^\prime = q_e^\prime$). 
Furthermore, reactor experiments~\cite{Beda:2010zz} give the bound for $q_{\nu}^\prime = q_e^\prime$ at masses above $500\,{\rm eV}$ (dashed green). Unlike the previous case, our projected sensitivity covers regions already excluded by other terrestrial bounds with the exception of a small region at masses $m_{\gamma^\prime} \simeq 1\, {\rm keV}$ if neutrino couplings are switched on. However, this region is excluded by the solar arguments and therefore it has to be carefully identified through evolutionary solar profiles as discussed already for the kinetic mixing scenario.

\section{Final Remarks}
\label{sec:Conclusion}

The last decade has been witnessing a regular increase in interest in light and weakly-coupled new particles. There are at least two reasons behind this activity: the lack of a new physics discovery at the weak scale by the Large Hadron Collider and searches for Weakly Interacting Massive Particles and the blossoming of several novel ideas for experimentally tackling such a low-mass region. Stars provide a natural environment to produce these hypothetical light states in their interior. In the Sun, thermal processes can produce particles up to masses around the keV scale, and supernovae can even go above the MeV scale. Stars like the Sun provide an additional non-thermal production process via nuclear reactions, allowing for the study of sub-MeV particles in these environments. Typically, cooling arguments put upper bounds on the couplings of these new particles that are often stronger than the ones obtained in laboratories.

In this work, we have employed the second step of the $pp$ chain in the solar core to produce \textit{``massive hidden vector bosons"}. Instead of relying upon cooling arguments, we have investigated the discovery reach of the forthcoming JUNO experiment to detect directly this monochromatic flux of 5.49 MeV spin-one particles. We focused on the Abelian extension of the SM gauge group, and we assumed the new gauge group to be broken spontaneously so that the resulting gauge boson field $A^\prime_\mu$ is massive. We restricted ourselves to vector-like dark gauge charges (if any) that are consistent with electroweak gauge invariance. This charge choice led to only isoscalar dark gauge interactions between nucleons and the new massive vector. Due to the isovector nature of the transition in the deuterium fusion process, tree-level dark gauge interactions cannot be responsible for the production of massive vectors. However, we have identified two possible mixing between the dark gauge boson and the SM photon that could control the production rate: the kinetic mixing between the two Abelian field strenghts~\cite{Holdom:1985ag}, and the plasma mixing through the thermal electron loop that is ensured once the electron has a non-vanishing dark gauge charge~\cite{Hardy:2016kme,Hong:2020bxo}. The production rates via these two mechanisms are shown in Figs.~\ref{fig:DPflux} and \ref{fig:BLflux}. Once these particles are produced and made their way to the Earth, they can be detected via Compton-like scatterings.

We have investigated the discovery potential of the JUNO detector by adopting the statistical analysis developed by Ref.~\cite{Lucente:2022esm}. The parameter space region within the reach of JUNO is shown in Fig.~\ref{Fig:JUNO_DP} and Fig.~\ref{Fig:JUNO_IM} for the kinetic mixing and plasma mixing scenarios, respectively. For the kinetic mixing scenario, JUNO can achieve a sensitivity better than current terrestrial bounds in the mass region $50 \, {\rm keV} \lesssim m_{\gamma^\prime} \lesssim {\rm MeV}$. On the contrary, for the plasma mixing scenario, JUNO will cover regions already excluded by other terrestrial bounds with the only exception of a tiny region around $m_{\gamma^\prime} \simeq {\rm keV}$. Finally, given the high energy threshold, it would be challenging for the upcoming neutrino oscillation experiment like Deep Underground Neutrino Experiment \cite{DUNE:2020ypp}  at Fermilab to detect 5.49 MeV hidden sector particles. On the other hand, its competing project Hyper-Kamiokande experiment \cite{Hyper-Kamiokande:2018ofw} in Japan has an appropriate energy threshold but because of its lower energy resolution, $\bar{\sigma}/{\rm MeV}=0.6 \sqrt{E/{\rm MeV}}$, provides an order of magnitude weaker bounds to hidden sectors than the JUNO (see Ref.~\cite{Lucente:2022esm} for bounds on axion couplings).

\begin{figure}[t!]
\centering
\includegraphics[width=0.65\textwidth]{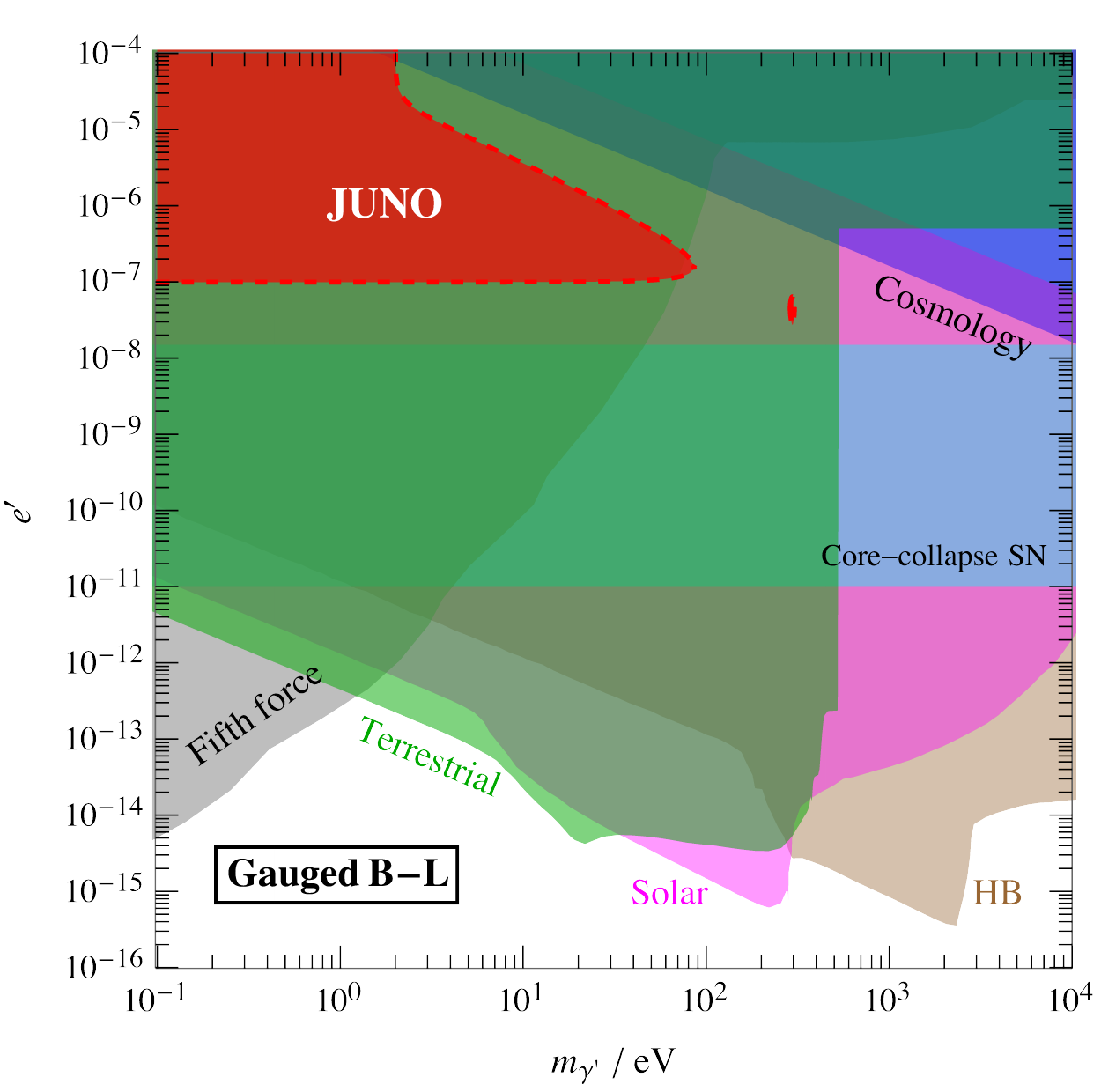} 
\caption{\footnotesize Exclusion region plot for the $B-L$ extension of the SM. The red region in the $(m_{\gamma^\prime}, e^\prime)$ plane (with $e^\prime$ the dark gauge coupling) represents the JUNO discovery reached the 90\% C.L. The blue, gray, and green regions indicate the current constraints based on cosmology, fifth force tests, and terrestrial experiments, respectively. The magenta, brown, and cyan regions are excluded by astrophysics with respect to the Sun, horizontal branch stars, and core-collapse supernovae, respectively.}
\label{Fig:JUNO_BL}
\end{figure}

The two scenarios studied in this work are useful for exemplifying the most general case but are certainly not exhaustive. We can easily extend the framework developed here to study any $U(1)$ gauge extensions of SM. A very instructive example is the gauged $B-L$ model (if we want the model to be anomaly-free we need to introduce right-handed neutrinos). The baryon number gauge charge assignments lead to isoscalar coupling and therefore negligible effects on the production of solar $\gamma^\prime$, the lepton number charge induces the plasma mixing for the production and mediates detection via scatterings and decays. The lepton number also involves neutrinos, so the solar $\gamma^\prime$ flux on the Earth would be further reduced by $\gamma^\prime \rightarrow \nu\bar{\nu}$ decays as discussed above. The kinetic mixing parameter in this case is a free parameter in the Lagrangian that has to be fixed by some renormalization condition. Even if we set it to vanish at some energy scale, it would be radiatively induced by loops of particles carrying both Abelian charges. If we assume the kinetic mixing to be vanishing at some UV scale $\Lambda_{\rm UV}$, and we find that loop of SM particles gives a low-energy value that results in $\varepsilon \simeq  (ee^\prime/3\pi^2)\log\left[\Lambda_{\rm UV}^2/v_{\rm EW}^2\right]$. Assuming $\Lambda_{\rm UV} = 100\,{\rm TeV}$ and taking $v_{\rm EW}\simeq 246\,{\rm GeV}$, we have the kinetic mixing ${\varepsilon \simeq 0.12\, e^\prime}$. Likewise, for the other motivated scenario where we gauge the combination of lepton numbers $L_{\mu}-L_\tau$, the irreducible kinetic mixing from radiative corrections results in $\varepsilon \simeq (e e^\prime/12\pi^2)\log (m_\tau^2/m_\mu^2) \simeq e^\prime / 70$. Fig.~\ref{Fig:JUNO_BL} shows the discovery reach for the $B-L$ case. The red region indicates the JUNO sensitivity at 90\%~C.L. (the small island around $\omega_{\rm pl}$ due to the resonant feature), and this result is valid regardless of the low-energy value of the kinetic mixing $\varepsilon$. In this case, neutrino decay depletes the signal in the parameter space region sensitive to kinetic mixing. These simple observations motivate the analysis of explicit UV complete models building upon the present work.

From a slightly broader perspective, our work contains novel results on the physics of massive spin-one particles that can be employed beyond the experimental setup of JUNO analyzed here. The production rates derived in Sec.~\ref{sec:production} quantify the incoming flux of feebly-interacting vectors reaching a generic detector on Earth. An option worthwhile to investigate is along the lines of the study performed by Ref.~\cite{Bhusal:2020bvx}, where dissociation of deuterons at the Sudbury Neutrino Observatory (SNO) was exploited as a probe of axion couplings to nuclei. The same analysis could be performed if the dissociation is generated by massive vectors produced in the solar core, and the production rates computed in this work will be an essential input to this analysis. On the opposite end of the story, the Compton-like scattering cross sections provided in Sec.~\ref{sec:detection} (with details in App.~\ref{app:Compton}) for polarized initial states could improve existing studies in the current literature. Whether we consider thermal production in the solar interior (with energy around the keV scale) or products of nuclear reactions (with energy around the MeV scale), the Sun is a powerful source of \textit{polarized} massive vectors. If we do not invoke any cooling argument but instead aim at detecting the feebly-interacting spin-one bosons on Earth, the proper procedure is to evaluate the detection cross sections for the different polarizations with the appropriate weight. This signal is actually searched for by dark matter direct detection experiments (see, e.g., Refs.~\cite{Majorana:2016hop,XMASS:2018pvs,SuperCDMS:2019jxx,GERDA:2020emj}). Consistent implementation of the polarized fluxes is still missing in the literature, and the cross-sections provided in this work will also serve as crucial input. We defer these further applications of our calculations to future work.


\paragraph{Acknowledgments.} The authors thank A. Caputo, N. Houston, and M. Pospelov for useful discussions and feedback on the manuscript. The work of FD and SY is supported by the research grant ``The Dark Universe: A Synergic Multi-messenger Approach'' number 2017X7X85K under the program PRIN 2017 funded by the Ministero dell'Istruzione, Universit\`a e della Ricerca (MIUR). The work of  G.L. is supported by the ``NAT-NET: Neutrino and Astroparticle Theory Network'' number 2017W4HA7S under the program PRIN 2017 funded by the Ministero dell'Istruzione, Universit\`a e della Ricerca (MIUR). All authors are supported by Istituto Nazionale di Fisica Nucleare (INFN) through the Theoretical Astroparticle Physics (TAsP) project. SY acknowledges the IBS Center for Theoretical Physics of the Universe (CTPU) and the APCTP-supported program ``Dark Matter as a Portal to New Physics 2023'' for their kind hospitality during the completion of this work. This article is based upon work from COST Action COSMIC WISPers CA21106, supported by COST (European Cooperation in Science and Technology). GL and SY thank the Galileo Galilei Institute for Theoretical Physics for their hospitality during the preparation of part of this work.

\appendix

\section{Compton-like scattering cross section for polarized hidden vectors}
\label{app:Compton}

The Compton-like scattering where an incoming massive vector from the Sun hits a target electron at rest, and leads to a final state with one photon and one electron, is a key process for the hidden vector detection in our analysis. Crucially, the flux of vectors produced in the solar interior is polarized, and as a consequence, we have to compute the Compton-like cross section for each given polarization. We provide in this Appendix the computational details for the scattering cross section. 

We use the following notation to describe the four momenta involved in the scattering
\begin{equation*}
\gamma^\prime(k) + e^-(p_1) \, \rightarrow \, \gamma(q) + e^-(p_2) \ .
\end{equation*}
Here, we put in parenthesis the four-momentum associated with each particle. Conservation of energy and momentum reads $k + p_1 = q + p_2$, and we find it useful to introduce the Mandelstam variables
\begin{align*}
s = & \, (k + p_1)^2 = (q + p_2)^2 \ , \\
t = & \, (p_1 - p_2)^2 = (q - k)^2 \ , \\
u = & \, (p_1 - q)^2 = (k - p_2)^2 \ .
\end{align*}
They are related by the constraint $s+t+u = 2 m_e^2 + m_{\gamma^\prime}^2$. 

\subsection*{Laboratory frame kinematics}

\begin{figure}[t!]
\centering
\includegraphics[width=0.95\textwidth]{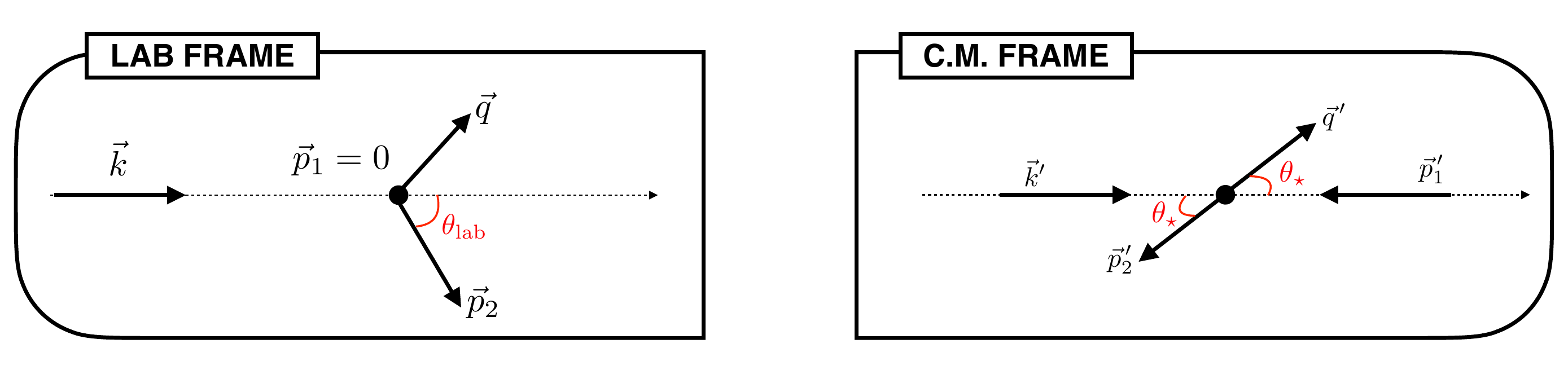}
\caption{\footnotesize Scattering angles in the laboratory frame where the initial state electron is at rest (left) and the center of mass frame (right).}
\label{fig:AnglesFrame}
\end{figure}

We perform the calculation in the laboratory frame where the target electron is at rest and the incoming vector is monochromatic. Initial state four-momenta read $k^\mu = (\omega, k \hat{k})$ and $p_1^\mu = (m_e, \vec{0})$ with $\omega$ fixed to 5.49 MeV and $k^2 = \omega^2 - m_{\gamma^\prime}^2$. The polarization vectors are indeed the ones given in Eqs.~\eqref{eq:epsilonT} and \eqref{eq:epsilonL}. We reproduce them here for convenience and this time we use the dispersion relation to write them only in terms of $\omega$ and $m_{\gamma^\prime}$
\begin{align}
\text{Transverse: } & \qquad \epsilon_{\rm T}^\mu =  \left(0, \frac{\hat{e}_1 \pm i \hat{e}_2}{\sqrt{2}} \right) \ , \\
\text{Longitudinal: } & \qquad \epsilon_{\rm L}^\mu =  \frac{1}{m_{\gamma^\prime}} \left(\sqrt{\omega^2 - m_{\gamma^\prime}^2}, \omega \hat{k} \right) \ .
\label{eq:epsapp}
\end{align}
As usual, $\hat{e}_1$ and $\hat{e}_2$ are two normal unit vectors in the orthogonal plane. We define the final state electron four-momentum as $p_2^\mu = (E_{\rm lab}, \vec{p}_{\rm lab})$ with $p_{\rm lab}^2 = E_{\rm lab}^2 - m_e^2$, and $\theta_{\rm lab}$ to be the angle that the electron spatial momentum makes with the direction of the incoming vector: $\vec{k} \cdot \vec{p}_{\rm lab} = | \vec{k} | \left| \vec{p}_{\rm lab} \right| \cos\theta_{\rm lab}$. With these definitions, the Mandelstam variables explicitly read 
\begin{align*}
s = & \, m_{\gamma^\prime}^2 + m_e^2 + 2 \omega m_e \ , \\ 
t = & \, - 2 m_e (E_{\rm lab} - m_e) \ , \\ 
u  = & \, - m_e (2 \omega - 2 E_{\rm lab} + m_e) \ .
\end{align*}
The laboratory frame kinematics is sketched in the left panel of Fig.~\ref{fig:AnglesFrame}. Our ultimate goal is to compute the scattering cross section. We notice how the Mandelstam variables are completely identified by one piece of information about the initial state, the value of $\omega$, and one piece of information about the final state, the electron energy $E_{\rm lab}$. In particular, they do not depend on the scattering angle $\theta_{\rm lab}$. If we were computing the unpolarized cross section, this set of information would be enough. However, we are going to compute the scattering cross section for a fixed polarization of the incoming vector (see Eq.~\eqref{eq:epsapp}) and therefore we have to compute Lorentz invariant product between the particle four-momenta and the polarization vectors themselves. We do not need to compute all possible combinations. First, we know how the polarization vectors are orthogonal to the four-vector of the particle they describe (i.e., $k \cdot \epsilon_{P \mu}(k) = 0$). Furthermore, we can always use four-momentum conservation to express $q \cdot \epsilon_{P \mu}(k)$ in terms of the other Lorentz invariant products. Ultimately, we only need to compute the Lorentz invariant products between the polarization vectors and the electron four-momenta
\begin{align}
\label{eq:pepsT} \text{Transverse: } & \, \qquad p_1 \cdot \epsilon_{\rm T} = 0  \ , \;  \\ & \, \qquad p_2 \cdot \epsilon_{\rm T} = - \vec{p}_{\rm lab} \cdot \frac{\hat{e}_1 \pm i \hat{e}_2}{\sqrt{2}} \ . \\
\label{eq:pepsL} \text{Longitudinal: } & \, \qquad p_1 \cdot \epsilon_{\rm L} = 
\frac{m_e \, \sqrt{\omega^2 - m_{\gamma^\prime}^2}}{m_{\gamma^\prime}}   \ , \; \\ & \, \qquad 
p_2 \cdot \epsilon_{\rm L} = \frac{E_{\rm lab} \sqrt{\omega^2 - m_{\gamma^\prime}^2} - \omega \sqrt{E_{\rm lab}^2 - m_e^2} \cos\theta_{\rm lab}}{m_{\gamma^\prime}} \ . 
\end{align}
The Lorentz invariant products between the initial state electron four-momentum and the polarization vectors are straightforward. The corresponding quantities with the final state electron four-momentum require more attention. First, we notice how the products with the two transverse polarization vectors in Eq.~\eqref{eq:pepsT} isolate the two components of $\vec{p}_{\rm lab}$ orthogonal to the incoming vector direction. The analogous quantity with the longitudinal polarization in Eq.~\eqref{eq:pepsL} depends on both $E_{\rm lab}$ and $\cos\theta_{\rm lab}$, and these two kinematical variables are not independent. We choose to identify the final state with the value of $E_{\rm lab}$. The connection between $\cos\theta_{\rm lab}$ and $E_{\rm lab}$ is eased once we study the kinematics in the center of mass frame. 

\subsection*{Center of mass kinematics} 

The center of mass frame (or, more properly, the center of momentum frame) is identified by the condition that the total spatial momentum is vanishing. We introduce a generic Lorentz boost connecting the two frames as follows
\bea
E^\prime = \gamma(\beta_\star) \left[E - \beta_\star p_{\parallel}  \right] \ ,  \qquad  \qquad 
p^\prime_{\parallel} = \gamma(\beta_\star) \left[p_{\parallel} - \beta_\star E \right] \ ,   \qquad \qquad 
p^\prime_{\perp} = p_{\perp} \ .
\eea
Primed and unprimed kinematical variables correspond to the center of mass and laboratory frames, respectively. Furthermore, we have the Lorentz factor $\gamma(\beta_\star) = \left(1 - \beta_\star^2 \right)^{-1/2}$, and the parallel ($\parallel$) and orthogonal ($\perp$) directions are identified with respect to the incoming vector spatial momentum. The requirement that the initial state momentum along the incoming direction ($p_\parallel$) is vanishing in the center of mass frame leads to the relations
\bea
\beta_\star = \frac{\sqrt{\omega^2 - m_{\gamma^\prime}^2}}{\omega + m_e} \ , \qquad \qquad \qquad
\gamma(\beta_\star) = \frac{\omega + m_e}{\sqrt{m_{\gamma^\prime}^2 + m_e^2 + 2 \omega m_e}} = \frac{\omega + m_e}{\sqrt{s}} \ .
\eea 
The advantage of working in the center of mass frame is that both initial and final states are monochromatic. In particular, the final state electron has energy and spatial momentum size that are function of $\sqrt{s}$ only
\begin{equation*}
E^\prime_2 = \frac{\sqrt{s}}{2} \left(1 + \frac{m_e^2}{s} \right) \ , \qquad \qquad \qquad
\left| \vec{p}^{\,\prime}_2 \right| = \frac{\sqrt{s}}{2} \left(1 - \frac{m_e^2}{s} \right) \ .
\end{equation*}
Thus the final state is identified by the scattering angle $\theta_\star$ that is illustrated in the left panel of Fig.~\ref{fig:AnglesFrame}. 

\subsection*{Final state electron kinematical variables.} 

The relations identified above in the two frames allow us to describe completely the final state electron. First, we employ the inverse Lorentz boost to write down the electron energy in the laboratory frame 
\bea
E_{\rm lab} = \gamma(\beta_\star) 
\left[E^\prime_2 - \beta_\star \left| \vec{p}^{\,\prime}_2 \right| \cos\theta_\star  \right] \ .
\label{eq:ElabvsCM}
\eea
The scattering angle in the center of mass frame can take any value between $0$ and $\pi$. Thus the final state electron energy is bounded from both above and below, $E_{\rm lab}^{-} \leq E_{\rm lab} \leq  E_{\rm lab}^{+}$. The kinematical thresholds explicitly read
\bea
E_{\rm lab}^{\pm} =  \frac{(\omega + m_e) (s + m_e^2) \pm \sqrt{\omega^2 - m_{\gamma^\prime}^2} (s - m_e^2)}{2 s} \ .
\label{eq:Elabvsthetastar}
\eea
The only leftover task is to express the scattering angle $\theta_{\rm lab}$ as a function of $E_{\rm lab}$. The center of mass frame kinematics is useful here as well
\bea
\tan\theta_{\rm lab} = \frac{p_{2 \perp}}{p_{2 \parallel}} = \frac{- \left| \vec{p}^{\,\prime}_2 \right| \sin\theta_\star}{\gamma(\beta_\star) \left[- \left| \vec{p}^{\,\prime}_2 \right| \cos\theta_\star + \beta_\star E_2^\prime \right]} \ .
\eea
We use the relation in Eq.~\eqref{eq:Elabvsthetastar} to express the center of mass scattering angle $\theta_\star$ in terms of $E_{\rm lab}$, and we take the energy and the size of the spatial momentum in the center of mass frame from Eq.~\eqref{eq:ElabvsCM}. Once we put everything together, and we convert the tangent into a cosine (since it is what we need, see Eq.~\eqref{eq:pepsL}), we find
\bea
\cos\theta_{\rm lab} = \frac{2 \left(m_e+\omega \right) \left(E_{\rm lab}- m_e \right) - m_{\gamma^\prime}^2}{2 \sqrt{\omega^2 - m_{\gamma^\prime}^2} \sqrt{E_{\rm lab}^2-m_e^2}} \ .
\eea

\subsection*{Scattering cross section} 

The differential cross section reads
\begin{equation*}
d \sigma_P = \frac{\overline{\left| \mathcal{M}_P \right|^2}}{4 I}  d \Phi_{\rm final} \ ,
\end{equation*}
where the subscript $P$ refers to the polarization state of the incoming vector that can be either transverse ($P={\rm T}$) or longitudinal ($P={\rm L}$). The Lorentz invariant flux factor $I$ appearing in the cross section results in
\begin{equation*}
I = \sqrt{\left(k \cdot p_1 \right)^2 - m_{\gamma^\prime}^2 m_e^2} = m_e \sqrt{\omega^2 - m_{\gamma^\prime}^2} \ .
\end{equation*}
The Lorentz-invariant phase space factor $d \Phi_{\rm final}$ accounts for all possible kinematically allowed final states. In the center of mass frame, the Lorentz invariant phase space takes the well-known form and once we use the information that the squared matrix element respects cylindrical symmetry around the incoming direction of the massive vector the phase space is just proportional to $d \cos\theta_\star$. The relation in Eq.~\eqref{eq:ElabvsCM} allows us to derive the corresponding quantity in the laboratory frame, and after we differentiate it we find
\begin{equation*}
d \Phi_{\rm final} = \frac{1}{16 \pi} \left(1 - \frac{m_e^2}{s} \right) d \cos\theta_\star = \frac{1}{8 \pi \sqrt{\omega^2 - m_{\gamma^\prime}^2}} d E_{\rm lab}  \ .
\end{equation*}
To summarize, the general expression for the differential cross section reads
\bea
\frac{d \sigma_P}{d E_{\rm lab}} = 
\frac{\overline{\left| \mathcal{M}_P \right|^2}}{32 \pi \, m_e (\omega^2 - m_{\gamma^\prime}^2)} \ . 
\label{eq:sigmadiff}
\eea

\begin{figure}[t!]
\centering
\includegraphics[width=0.95\textwidth]{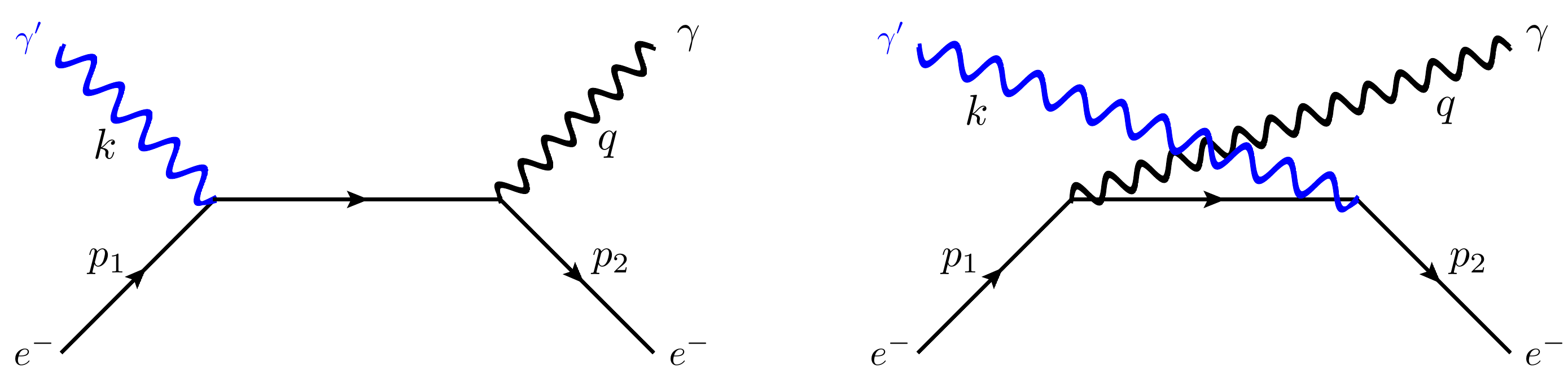}
\caption{\footnotesize Feynman diagrams for the Compton-like scattering. The blue lines correspond to the hidden vector $\gamma^\prime$ whereas regular wavy black lines to the SM photon $\gamma$. We have contributions from both the s-channel (left) and the u-channel (right).}
\label{fig:ComptonFeynman}
\end{figure}

\subsection*{Polarized squared matrix elements} 

The couplings in the Lagrangian that mediate this process are 
\begin{equation*}
\mathcal{L}_{\rm Compton-like} = \left( - e A_\mu + g_e A^\prime_\mu \right) \, \bar{e} \gamma^\mu e \ .
\end{equation*}
The first term comes from the electromagnetic interaction of the electron field, the coupling with the dark vector field has to account for mixing effects as well. As discussed in this paper, the general expression reads $g_e = - \varepsilon e + q_e^\prime e^\prime$. We have a polarized flux of dark vectors coming from the Sun, and for this reason, we evaluate the transition amplitude for a fixed initial-state polarization $P$ of the incoming $\gamma^\prime$ particle. The Feynman diagrams for the Compton-like scattering are shown in Fig.~\ref{fig:ComptonFeynman}, and we write the total matrix element as the sum of these two contributions
\begin{equation*}
i \, \mathcal{M}_P = i \, \mathcal{M}^{(s)}_P + i \, \mathcal{M}^{(u)}_P \ .
\end{equation*}
The two individual terms contributing to the transition amplitude, s- and u-channels, explicitly read
\begin{align}
i \, \mathcal{M}^{(s)}_P = & \, \bar{u}(p_2) (- i e \gamma^\nu) 
\frac{i(\slashed{p}_1 + \slashed{k} + m_e)}{s - m_e^2} (i g_e \gamma^\mu) u(p_1) \;\; \epsilon_{P \mu}(k)  \, \epsilon^\gamma_{i \nu}(q)^* \ , \\ 
i \, \mathcal{M}^{(u)}_P = & \, \bar{u}(p_2) (i g_e \gamma^\mu) 
\frac{i(\slashed{p}_1 - \slashed{q} + m_e)}{u - m_e^2} (- i e \gamma^\nu) u(p_1) \;\; \epsilon_{P \mu}(k) \, \epsilon^\gamma_{i \nu}(q)^* \ .
\end{align}
The dark gauge boson polarization vectors $\epsilon_{P \mu}(k)$ are given in Eqs.~\eqref{eq:epsilonT} and \eqref{eq:epsilonL}. We sum over the final state photon polarizations $\epsilon^\gamma_{i \nu}(q)$, and only the two transverse ones given in Eq.~\eqref{eq:epsilonT} are available in this case. Once we square the matrix element, and average and sum over initial and final states with the only exception of the fixed polarization $P$, we find the general expression
\bea
\overline{\left| \mathcal{M}_P \right|^2} = \frac{e^2 g_e^2}{2} \, L^{\mu\nu\alpha\beta} \
\left(\sum_{i = \pm} \epsilon^\gamma_{i \beta}(q) \epsilon^\gamma_{i \nu}(q)^* \right) \, \left(\epsilon_{P \mu}(k) \epsilon_{P \alpha}(k)^* \right)  \ .
\eea
The overall factor of $1/2$ averages over the initial state electron spins. The leptonic tensor explicitly reads
\bea
L^{\mu\nu\alpha\beta} = {\rm Tr} & \left[(\slashed{p}_2 + m_e) \left( \gamma^\nu \frac{\slashed{p}_1 + \slashed{k} + m_e}{s - m_e^2} \gamma^\mu + \gamma^\mu \frac{\slashed{p}_1 - \slashed{q} + m_e}{u - m_e^2} \gamma^\nu \right) \times \right. \nonumber  \\ & \left.  (\slashed{p}_1 + m_e) 
\left( \gamma^\alpha \frac{\slashed{p}_1 + \slashed{k} + m_e}{s - m_e^2} \gamma^\beta + \gamma^\beta \frac{\slashed{p}_1 - \slashed{q} + m_e}{u - m_e^2} \gamma^\alpha \right)\right] \ .
\eea
We always sum over the final-state photon polarizations
\bea
\sum_{i = \pm} \epsilon^\gamma_{i \beta}(q) \epsilon^\gamma_{i \nu}(q)^* = - g_{\nu\beta} \ .
\eea
As anticipated, we do not average over the initial state polarizations $P$ of the massive vector, and this makes the calculation more difficult since we cannot employ the well-known result for the sum over {\it all} polarizations
\bea
\sum_P \epsilon_{P \mu}(k) \epsilon_{P \nu}(k)^* = 
\sum_{T = \pm} \epsilon_{T \mu}(k) \epsilon_{T \nu}(k)^* + 
\epsilon_{L \mu}(k) \epsilon_{L \nu}(k)^* = - g_{\mu\nu} + \frac{k_\mu k_\nu}{m_{\gamma^\prime}^2} \ .
\label{eq:sumallpol}
\eea
We work in the laboratory frame where the polarization vectors are defined. For a generic laboratory frame that moves with four-velocity $U_\mu$, we have the following relations~\cite{Weldon:1996kb}
\begin{align}
\sum_{T = \pm} \epsilon_{T \mu}(k) \epsilon_{T \nu}(k)^* = & \, 
- \tilde{g}_{\mu\nu} - \frac{\tilde{k}_{\mu} \tilde{k}_{\nu}}{k^2} \, , \\
\epsilon_{L \mu}(k) \epsilon_{L \nu}(k)^* = & \, - g_{\mu\nu} + \frac{k_\mu k_\nu}{m_{\gamma^\prime}^2} - \left( - \tilde{g}_{\mu\nu} - \frac{\tilde{k}_{\mu} \tilde{k}_{\nu}}{k^2} \right) \ ,
\end{align}
where we define $\tilde{g}_{\mu\nu} = g_{\mu\nu} - U_\mu U_\nu$, and $\tilde{k}_{\mu} = k_{\mu} - (k \cdot U)U_\mu$. For our specific calculation, the target electron is at rest and therefore $U_\mu = (1\,,\vec{0})$. Thus the Lorentz boost to achieve the center of mass frame is aligned with the momentum of the massive vector. Furthermore, we take advantage of the fact that the hidden vector couples to a conserved current and therefore if we write the matrix element $\mathcal{M}_P = \epsilon_{P \mu}(k) M^\mu$ the condition $k_\mu M^\mu = 0$ must be satisfied. This has two immediate consequences. First, the sum over all polarizations can be replaced by Eq.~\eqref{eq:sumallpol} without the last term proportional to $k_\mu k_\nu$. Second, the squared matrix element for the longitudinal vector takes a particularly simple form after we plug the explicit expression for the polarization vector and we impose current conservation. In the end, the polarization part of the squared matrix element can be replaced by the following simple expressions~\cite{Shin:2022ulh}
\begin{align}
\label{eq:concurT} \sum_{T = \pm} \epsilon_{T \mu}(k) \epsilon_{T \nu}(k)^* \xrightarrow[]{k_\mu M^\mu = 0}  & \, 
- g_{\mu\nu} - \frac{m_{\gamma^\prime}^2}{\omega^2 - m_{\gamma^\prime}^2}g_{\mu 0}g_{\nu 0} \, , \\
\label{eq:concurL} \epsilon_{L \mu}(k) \epsilon_{L \nu}(k)^* \xrightarrow[]{k_\mu M^\mu = 0} & \, \frac{m_{\gamma^\prime}^2}{\omega^2 - m_{\gamma^\prime}^2}g_{\mu 0}g_{\nu 0} \ .
\end{align}
For a generic case where the massive vector couples to a non-conserved current (e.g., the electron axial current), we have to include the explicit expression for the polarization vectors. As already discussed, we can always employ four-momentum conservation and the conditions $k^\mu \epsilon_{P \mu}(k) = 0$ and  $\epsilon_{P \mu}(k) \epsilon_{P}^\mu(k) = - 1$ to have only the Lorentz invariant with electron initial and final state momenta provided by Eqs.~\eqref{eq:pepsT} and \eqref{eq:pepsL}. We performed our calculations by using both methods, conditions in Eqs.~\eqref{eq:concurT} and \eqref{eq:concurL} from current conservation and explicit polarization vectors, and we employed the {\tt FeynCalc} package~\cite{Shtabovenko:2016sxi,Shtabovenko:2020gxv} to perform the Dirac algebra and the Lorentz indices contractions. In the next subsection, we provide our final results for both the differential and total cross sections.
 
\subsection*{Final expressions for the differential and total cross sections} 

Once we have the squared matrix elements, we can write down the differential cross section as given by Eq.~\eqref{eq:sigmadiff}. For the $P={\rm T}$ case, we sum over the two transverse polarizations and average over them, whereas we have one only one option for the $P={\rm L}$ case. We find the differential scattering cross sections 
{\tiny \begin{dmath}
\frac{d \sigma_{\rm T}}{d E_{\rm lab}} = \frac{e^2 g_e^2}{64 \pi  m_e^2 (\omega^2-m_{\gamma^\prime}^2)^2 (2 \omega m_e + m_{\gamma^\prime}^2)^2 (m_e - E_{\rm lab} + \omega)^2} \, \times \, \left[ 8 \omega  m_e^6 \left(-4 \omega  E_{\rm lab}+m_{\gamma^\prime}^2+6 \omega ^2\right)-2 m_e m_{\gamma^\prime}^4 \left(m_{\gamma^\prime}^2 \left(\omega -2 E_{\rm lab}\right) \left(E_{\rm lab}+\omega \right)-2 \omega 
   \left(\omega -E_{\rm lab}\right) \left(-6 \omega  E_{\rm lab}+4 E_{\rm lab}^2+3 \omega ^2\right)+m_{\gamma^\prime}^4\right)+8 m_e^5 \left(\omega  m_{\gamma^\prime}^2 \left(3 \omega -E_{\rm lab}\right)+2 \omega
   ^2 E_{\rm lab} \left(E_{\rm lab}-7 \omega \right)-m_{\gamma^\prime}^4+10 \omega ^4\right)+4 m_e^3 \left(m_{\gamma^\prime}^4 \left(2 E_{\rm lab} \left(\omega -2 E_{\rm lab}\right)-3 \omega ^2\right)+2 \omega 
   m_{\gamma^\prime}^2 \left(-16 \omega ^2 E_{\rm lab}+11 \omega  E_{\rm lab}^2+E_{\rm lab}^3+6 \omega ^3\right)+4 \omega ^3 \left(\omega -E_{\rm lab}\right) \left(-2 \omega 
   E_{\rm lab}+E_{\rm lab}^2+2 \omega ^2\right)\right)+m_e^2 \left(8 \omega ^2 m_{\gamma^\prime}^2 \left(\omega -E_{\rm lab}\right) \left(3 E_{\rm lab} \left(E_{\rm lab}-2 \omega \right)+4 \omega
   ^2\right)+4 m_{\gamma^\prime}^4 \left(-6 \omega ^2 E_{\rm lab}+7 \omega  E_{\rm lab}^2+E_{\rm lab}^3+\omega ^3\right)-6 \omega  m_{\gamma^\prime}^6\right)+4 m_e^4 \left(2 \omega  m_{\gamma^\prime}^2 \left(7 \omega
   ^2-E_{\rm lab} \left(E_{\rm lab}+11 \omega \right)\right)+5 m_{\gamma^\prime}^4 \left(E_{\rm lab}-\omega \right)+4 \omega ^3 E_{\rm lab} \left(5 E_{\rm lab}-8 \omega \right)+16 \omega ^5\right)+16
   \omega ^2 m_e^7-m_{\gamma^\prime}^6 \left(\omega -E_{\rm lab}\right) \left(m_{\gamma^\prime}^2-2 \left(2 E_{\rm lab} \left(E_{\rm lab}-\omega \right)+\omega ^2\right)\right) \right]
\end{dmath}
\begin{dmath}
\frac{d \sigma_{\rm L}}{d E_{\rm lab}} = \frac{e^2 g_e^2 m_{\gamma^\prime}^2 \left(E_{\rm lab} \left(2 \omega  m_e+m_{\gamma^\prime}^2\right)-\left(2 m_e+\omega \right) \left(2 m_e \left(m_e+\omega
   \right)+m_{\gamma^\prime}^2\right)\right)}{32 \pi  m_e^2 (m_{\gamma^\prime}^2-\omega ^2)^2 
   (2 \omega  m_e+m_{\gamma^\prime}^2)^2
   \left(m_e-E_{\rm lab}+\omega \right){}^2} \, \times \,  \left[ 4 E_{\rm lab}^2 \left(2 \omega  m_e+m_e^2+m_{\gamma^\prime}^2\right)-4 E_{\rm lab} \left(m_e+\omega \right) \left(2 m_e \left(m_e+\omega \right)+m_{\gamma^\prime 
   }^2\right)+4 \omega  m_e m_{\gamma^\prime}^2+8 \omega ^2 m_e^2+8 \omega  m_e^3+4 m_e^4+m_{\gamma^\prime}^4 \right] \ .
\end{dmath}}
The total cross section for transverse ($P = {\rm T}$) and longitudinal ($P = {\rm L}$) results from the integration over the kinematical extremes provided by Eq.~\eqref{eq:Elabvsthetastar}, and we find
\bea
\sigma_{P} = \int_{E_{\rm lab}^-}^{E_{\rm lab}^+} \frac{d \sigma_P}{d E_{\rm lab}} \ .
\eea
The explicit integrations lead to the following results
{\tiny \begin{dmath}
\sigma_{\rm T} = \frac{e^2 g_e^2}{128 \pi  m_e^2 \left(m_{\gamma^\prime}^2-\omega ^2\right){}^2 \left(2 \omega  m_e+m_{\gamma^\prime}^2\right)} \, \times \, \left[ -\frac{\left(m_{\gamma^\prime}^2+2 \omega  m_e\right){}^2 \left(m_{\gamma^\prime}^4+\left(2 \omega -m_e\right) m_e m_{\gamma^\prime}^2+2 \omega ^2 m_e^2\right) \left(\omega
   +m_e-\sqrt{\omega ^2-m_{\gamma^\prime}^2}\right){}^2}{\left(m_e^2+2 \omega  m_e+m_{\gamma^\prime}^2\right){}^2}+\frac{\left(m_{\gamma^\prime}^2+2 \omega  m_e\right){}^2
   \left(m_{\gamma^\prime}^4+\left(2 \omega -m_e\right) m_e m_{\gamma^\prime}^2+2 \omega ^2 m_e^2\right) \left(\omega +m_e+\sqrt{\omega ^2-m_{\gamma
   }^2}\right){}^2}{\left(m_e^2+2 \omega  m_e+m_{\gamma^\prime}^2\right){}^2}+\frac{4 \left(-\left(\left(\omega +2 m_e\right) m_{\gamma^\prime}^6\right)-m_e \left(2 \omega
   ^2+2 m_e \omega +m_e^2\right) m_{\gamma^\prime}^4+2 \omega  m_e^3 \left(5 \omega +2 m_e\right) m_{\gamma^\prime}^2+4 \omega ^2 m_e^4 \left(2 \omega +m_e\right)\right)
   \left(-\omega -m_e+\sqrt{\omega ^2-m_{\gamma^\prime}^2}\right)}{m_e^2+2 \omega  m_e+m_{\gamma^\prime}^2}+\frac{4 \left(-\left(\left(\omega +2 m_e\right) m_{\gamma
   }^6\right)-m_e \left(2 \omega ^2+2 m_e \omega +m_e^2\right) m_{\gamma^\prime}^4+2 \omega  m_e^3 \left(5 \omega +2 m_e\right) m_{\gamma^\prime}^2+4 \omega ^2 m_e^4 \left(2
   \omega +m_e\right)\right) \left(\omega +m_e+\sqrt{\omega ^2-m_{\gamma^\prime}^2}\right)}{m_e^2+2 \omega  m_e+m_{\gamma^\prime}^2}+2 \left(\log \left(\frac{\left(m_{\gamma
   }^2+2 \omega  m_e\right) \left(\omega +m_e-\sqrt{\omega ^2-m_{\gamma^\prime}^2}\right)}{m_e^2+2 \omega  m_e+m_{\gamma^\prime}^2}\right)-\log (2)\right) \left(m_{\gamma
   }^6-2 \omega ^2 m_{\gamma^\prime}^4+16 \omega ^2 m_e^4+16 \omega  m_e^3 \left(\omega ^2+m_{\gamma^\prime}^2\right)-4 m_e^2 \left(2 \omega ^4-7 m_{\gamma^\prime}^2 \omega
   ^2+m_{\gamma^\prime}^4\right)\right)-2 \log \left(\frac{\left(m_{\gamma^\prime}^2+2 \omega  m_e\right) \left(\omega +m_e+\sqrt{\omega ^2-m_{\gamma^\prime}^2}\right)}{2
   \left(m_e^2+2 \omega  m_e+m_{\gamma^\prime}^2\right)}\right) \left(m_{\gamma^\prime}^6-2 \omega ^2 m_{\gamma^\prime}^4+16 \omega ^2 m_e^4+16 \omega  m_e^3 \left(\omega
   ^2+m_{\gamma^\prime}^2\right)-4 m_e^2 \left(2 \omega ^4-7 m_{\gamma^\prime}^2 \omega ^2+m_{\gamma^\prime}^4\right)\right)-\frac{4 m_e \left(m_e^2+2 \omega  m_e+m_{\gamma
   }^2\right) \left(-m_{\gamma^\prime}^4+2 \omega  \left(\omega +2 m_e\right) m_{\gamma^\prime}^2+4 \omega ^2 m_e^2\right)}{\omega +m_e+\sqrt{\omega ^2-m_{\gamma
   }^2}}+\frac{4 m_e \left(m_e^2+2 \omega  m_e+m_{\gamma^\prime}^2\right) \left(-m_{\gamma^\prime}^4+2 \omega  \left(\omega +2 m_e\right) m_{\gamma^\prime}^2+4 \omega ^2
   m_e^2\right)}{\omega +m_e-\sqrt{\omega ^2-m_{\gamma^\prime}^2}} \right] \ .
   \label{eq:sigmaT}
\end{dmath}
\begin{dmath}
\sigma_{\rm L} = \frac{e^2 g_e^2 m_{\gamma^\prime}^2}{32 \pi  m_e^2 \left(m_{\gamma^\prime}-\omega \right){}^2 \left(m_{\gamma^\prime}+\omega \right){}^2 \left(2 \omega  m_e+m_{\gamma^\prime}^2\right)} \, \times \,  \left[ \frac{2 \sqrt{\omega ^2-m_{\gamma^\prime}^2} \left(4 m_e m_{\gamma^\prime}^2 \left(\omega ^2-5 m_e \left(m_e+\omega \right)\right)+m_{\gamma^\prime}^4 \left(\omega -3 m_e\right)+4
   m_e^2 \left(m_e+\omega \right) \left(\omega ^2-4 m_e \left(m_e+2 \omega \right)\right)\right)}{2 \omega  m_e+m_e^2+m_{\gamma^\prime}^2}+\left(-4 m_e^2 m_{\gamma
   }^2-4 m_e^2 \left(2 m_e+\omega \right) \left(2 m_e+3 \omega \right)+m_{\gamma^\prime}^4\right) \left(\log \left(-\frac{\left(2 \omega  m_e+m_{\gamma^\prime}^2\right)
   \left(m_e-\sqrt{\omega ^2-m_{\gamma^\prime}^2}+\omega \right)}{2 \omega  m_e+m_e^2+m_{\gamma^\prime}^2}\right)-\log \left(-\frac{\left(2 \omega  m_e+m_{\gamma^\prime}^2\right)
   \left(m_e+\sqrt{\omega ^2-m_{\gamma^\prime}^2}+\omega \right)}{2 \omega  m_e+m_e^2+m_{\gamma^\prime}^2}\right)\right) \right] \ .
   \label{eq:sigmaL}
\end{dmath}}
These expressions take the simple form in the limit of small hidden vector mass
\begin{align}
\sigma_T \simeq & \, \frac{e^2 g_e^2}{16 \pi m_e^2}
\frac{2 x_\omega (x_\omega (x_\omega+1) (x_\omega+8)+2)+((x_\omega-2) x_\omega-2) (2 x_\omega+1)^2 \log (2 x_\omega+1)}{x_\omega^3 \left(2 x_\omega + 1\right){}^2} \ .
\end{align}
\begin{align}
\sigma_L \simeq & \, \frac{e^2 g_e^2 m_{\gamma^\prime}^2}{16 \pi  m_e^4 } 
\frac{\left(\frac{2 x_{\omega } \left(x_{\omega }+1\right) \left(\left(x_{\omega }-8\right) x_{\omega }-4\right)}{2 x_{\omega}+1}+\left(x_{\omega }+2\right) \left(3 x_{\omega }+2\right) \log \left(2 x_{\omega }+1\right)\right)}{x_{\omega }^5}  \ ,
\end{align}
where we provide the expressions in terms of the dimensionless variable $x_\omega = \omega / m_e$ quantifying the initial state energy in units of the electron mass.

\bibliographystyle{JHEP}
\bibliography{VectorsJUNO}

\end{document}